\documentclass{amsart}
\usepackage{mathrsfs}
\usepackage{graphicx}
\usepackage{amsmath}
\usepackage{amscd}
\usepackage{amssymb}
\usepackage[all]{xy}
\usepackage{appendix}

\input xy
\xyoption{all}

\newcommand{\pd}{\partial}
\newcommand{\bC}{{\mathbb C}}
 \newcommand{\bH}{{\mathbb H}}

\newcommand{\bT}{\mathbf{T}}

\newcommand{\bR}{{\mathbb R}}
\newcommand{\bZ}{{\mathbb Z}}

\newcommand{\cO}{{\mathcal O}}
\newcommand{\cP}{{\mathcal P}}
 \newcommand{\cS}{{\mathcal S}}

\newcommand{\half}{\frac{1}{2}}

\newcommand{\vac}{|0\rangle}

\DeclareMathOperator{\Aut}{Aut}

\DeclareMathOperator{\Rea}{Re} \DeclareMathOperator{\tr}{tr}
\DeclareMathOperator{\Img}{Im}

\newtheorem{thm}{Theorem}[section]
\newtheorem{theorem/definition}{Theorem/Definition}[section]

\theoremstyle{remark}
\newtheorem{remark}{Remark}[section]

\theoremstyle{definition}

\newcommand{\be}{\begin{equation}}
\newcommand{\ee}{\end{equation}}
\newcommand{\bea}{\begin{eqnarray}}
\newcommand{\ben}{\begin{eqnarray*}}
\newcommand{\een}{\end{eqnarray*}}
\newcommand{\eea}{\end{eqnarray}}
\newcommand{\bet}{\begin{equation}
\begin{split}}
\newcommand{\eet}{\end{split}
\end{equation}}

\usepackage{color}

\definecolor{yellow}{rgb}{1,1,0}
\definecolor{orange}{rgb}{1,.7,0}
\definecolor{red}{rgb}{1,0,0} \definecolor{green}{rgb}{0,1,1}
\definecolor{white}{rgb}{1,1,1}

\definecolor{A}{rgb}{.75,1,.75}

\newcommand{\corr}[1]{\langle {#1} \rangle}

\begin{document}

\title
{Hermitian One-Matrix Model and KP Hierarchy}

\author{Jian Zhou}
\address{Department of Mathematical Sciences\\
Tsinghua University\\Beijing, 100084, China}
\email{jzhou@math.tsinghua.edu.cn}

\begin{abstract}
The partition functions of Hermitian one-matrix models are known to be
tau-functions of the KP hierarchy.
In this paper we explicitly compute the elements in Sato grassmannian
these tau-functions correspond to,
and use  them to compute the $n$-point functions of Hermitian one-matrix model.
\end{abstract}
\maketitle

\section{Introduction}

Matrix models have been extensively studied by physicists and mathematicians
because of its connections to many different fields and the numerous
methods used to study them.
See e.g. \cite{Mehta} for a comprehensive introduction for their history
and various developments. See \cite{BKSWZ} for introduction to various connections to
other branches of mathematics and physics,
including Riemann zeta function, 2D quantum gravity, topological strings, moduli
spaces of Riemann surfaces, conformal field theory, etc.
For more comprehensive surveys on some of these topics,
see e.g. \cite{DGZ} and \cite{Marino}.
For applications to the moduli spaces of algebraic curves,
see e.g. \cite{Har-Zag, Pen, Wit1,Wit2, Kon}.

An important problem in the study of matrix models is to compute their
$n$-point correlations functions.
One approach is first to recursively solve
the loop equations
(see e.g. \cite{ACKM}).
This approach has been developed into the Eynard-Orantin topological
recursion \cite{Eyn-Ora}.
Another approach is to apply the connection to integrable hierarchies.
Since we are interested in the finite $N$ case,
the following two results are most relevant.
First of all,
the partition functions $Z_N$ of Hermitian one-matrix models for all $N \geq 1$
form a tau-function of   the Toda lattice hierarchy  \cite{UT}.
See e.g. \cite{GMMMO}.
This fact is recently used by Dubrovin and Yang \cite{Dub-Yan} to develop a formula
for the  $n$-point functions.
Secondly,
for each $N$, $Z_N$ is a tau-function
of the KP hierarchy.
See \cite{Shaw-Tu-Yen, Mulase}.
This is the point of departure of this paper.
Since we use a different integrable hierarchy from that used by
Dubrovin-Yang \cite{Dub-Yan},
our formula is different from theirs.

In an early work \cite{Zhou},
for any tau-function of the KP hierarchy given by a formal
power series,
we have proved a formula for its associated $n$-point functions.
This is based on Kyoto school's approach to KP hierarachy:
Sato grassmannian and boson-fermion correspondence.
This formula will be recalled in \S \ref{sec:Formula}.
To apply it,
one needs to convert the partition function which is originally an expression
involving Newton power functions
into an expression involving Schur functions.
In general this is not an easy task.
Fortunately in this case the partition function can be related to
the representation theory of symmetric groups \cite{Itz-Zub} (cf. \S \ref{sec:Itz-Zub}),
and also to the representation theory of unitary groups \cite{Marino} (cf. \S \ref{sec:Gro-Tay}).
As a result,
we get explicit expressions for the element in the Sato
grassmannian to apply our formula for $n$-point functions.

The partition functions $Z_N$ gives a family of tau-functions of the KP hierarchy,
one for each positive integer $N$.
This can be made into a continuous family $Z_t$ by introducing
the 't Hooft coupling constant.
Our formula for $N$-point functions also works for this family,
by replacing $N$ by $t$.
This will be explained in \S \ref{sec:tHooft}.

KP hierarchy and representation theory also appear in the study
of some geometric objects such as Hurwitz numbers,
Mari\~no-Vafa formula and open string invariants of conifolds.
See e.g. \cite{Zhou1}.
We will report on their corresponding elements in Sato grassmannian and
formulas for their $n$-point functions in a separate paper.

We arrange the rest of the paper in the following way.
In \S \ref{sec:Prelim} we recall some preliminaries of Hermitian matrix model,
including the computations of correlators by fat graphs
and by representation theory of symmetric groups.
In \S \ref{sec:Sato} we explicitly
compute the elements in Sato grassmannian that correspond to $Z_N$,
and use them to compute the $n$-point functions.
Our main result is formulated in Theorem \ref{thm:Main} and Theorem \ref{thm:Main2}.
In an Appendix we present some concrete computational results for the $n$-point functions.

\section{Preliminaries of Hermitian One-Matrix Model}

\label{sec:Prelim}

In this Section we recall the computations in Hermitian matrix model
by fat graphs and by representation theory of symmetric groups.

\subsection{Formal quantum field theory}

Let us recall the notion of a formal quantum field theory
introduced in \cite{Zhou2}.
It consists of {\em observable algebra} and {\em correlation functions}.
By an {\em observable algebra} we mean a commutative algebra $\cO$ with identity $1$,
whose elements are referred to as the observables.
In our examples,
an observable algebra is often an algebra of operators acting on
some space of functions.
The correlation functions are a sequence of homogeneous polynomial functions
on $\cO$ with values in a commutative algebra $R$.
When $\cO$ is generated by $\{\cO_i\}_{i\geq 0}$,
where $\cO_0 = 1$,
and
$$\cO_m \cO_n = \cO_n \cO_m,$$
for all $m, n\geq 0$,
then the correlation functions are  are determined by
the correlators $\corr{\cO_{m_1}, \dots, \cO_{m_n}} \in R$.
The {\em normalized correlators} $\corr{\cO_{m_1}, \dots, \cO_{m_n}}'$
are defined by
\be
\corr{\cO_{m_1}, \dots, \cO_{m_n}}':= \frac{\corr{\cO_{m_1}, \dots, \cO_{m_n}}}{\corr{\cO_0}}.
\ee
The {\em partition function} is defined by:
\be
Z:= 1+\sum_{n \geq 1} \corr{\cO_{m_1}, \dots, \cO_{m_n}}' \frac{t_{m_1} \cdots t_{m_n}}{n!},
\ee
where $\{t_0, t_1, \dots, t_n, \dots\}$ are formal variables.
The {\em free energy} $F$ is defined by:
\be \label{eqn:F-Z}
F:= \log Z.
\ee
The {\em connected correlators} are defined by:
\be \label{eqn:corr-F}
\corr{\cO_{m_1}, \dots, \cO_{m_n}}'_c
= \frac{\pd^nF}{\pd t_{m_1} \cdots \pd t_{m_n}} \biggl|_{t_i=0, i \geq 0}.
\ee
The $n$-point part of the free energy is defined by:
\be
F^{(n)}: = \frac{1}{n!} \sum_{m_1, \ldots, m_n}
\corr{\cO_{m_1}, \dots, \cO_{m_n}}'_c t_{m_1}\cdots t_{m_n}.
\ee

\subsection{Hermitian one matrix model as a formal quantum field theory}

We now explain that  the Hermitian one matrix model can be regarded
as a formal quantum  field theory.
We  take the observable algebra to be the algebra $\Lambda$ of symmetric functions.
We follow the notations in \cite{Macdonald}.
Denote by $\bH_N$ the space of Hermitian $N \times N$ matrices.
The algebra $\cS_N$ of $U(N)$-invariant polynomial functions on
the space $\bH_N$ is generated by
$M \to \tr (M^n)$.
We let $\Lambda$ act on $\cS_N$ as follows:
For $g \in \cS_N$,
define
\be
(p_n \cdot g ) (M) : = g_s^{-1}\tr(M^n) \cdot g(M).
\ee

For correlator we  use Gaussian integrals on $\bH_N$.
On this space consider the Euclidean measure:
\be
dM = 2^{N(N-1)/2} \prod^N_{i=1} dM_{ii} \prod_{1\leq i<j \leq N}
d\Rea(M_{ij})d\Img (M_{ij}).
\ee
For a polynomial function $f(M)$ in $M_{ij}$'s,
define
\be
\corr{f(M)}_N :
= \frac{\int_{\bH_N} dM f(M) e^{-\frac{1}{2g_s}\tr (M^2)}}{\int_{\bH_N} dM
e^{-\frac{1}{2g_s}\tr (M^2)}}.
\ee
For a partition $\lambda=(\lambda_1 \geq \lambda_2 \geq \cdots \geq \lambda_l)$,
define
\be
\corr{p_\lambda}_N:= g_s^{-l} \corr{\tr(M^{\lambda_1}) \cdots \tr(M^{\lambda_l})}_N.
\ee
Introduce formal variables $g_1, g_2, \dots$,
and let $g_\lambda=g_{\lambda_1} \cdots g_{\lambda_l}$.
As usual,
when $\lambda$ is the empty partition $\emptyset$,
$p_\emptyset = 1$ and $g_\emptyset = 1$.
Define the partition function by
\be
Z_N:= \sum_{\lambda \in \cP} \frac{1}{z_\lambda} \corr{p_\lambda}_N g_\lambda,
\ee
where the summation is taken over the set $\cP$ of all partitions.
It can be formally written as follows:
\be
Z_N = \frac{\int_{\bH_N} dM \exp \biggl( \tr
\sum\limits_{n=1}^\infty \frac{g_n-\delta_{n,2}}{ng_s} M^n\biggr)}{\int_{\bH_N} dM \exp \biggl( -\frac{1}{2g_s} \tr (M^2)\biggr)}.
\ee

\subsection{Feynman expansion for Hermitian one-matrix model}
\label{sec:Feynman}

The correlators $\frac{1}{z_\lambda}\corr{p_\lambda}_N$ can be evaluated by
the fat graph introduced by t'Hooft \cite{tH} based on Wick's formula
for Gaussian type integrals.
See \cite{BIZ} for an exposition.
One can check that:
\be
\corr{M_{ij}}_N = 0, \;\;\;\;
\corr{M_{ij}M_{kl}}_N = g_s\delta_{il}\delta_{jk}.
\ee
In general, Wick's formula gives for correlators of odd degree monomials
\ben
\corr{M_{i_1j_1}M_{i_2j_2}\cdots M_{i_{2n-1}j_{2n-1}}}_N = 0,
\een
and for corellators of even degree monomials:
\ben
&& \corr{M_{i_1j_1}M_{i_2j_2}\cdots M_{i_{2n}j_{2n}}}_N  \\
& = & \frac{1}{2^nn!} \sum_{\sigma \in S_{2n}}
\corr{M_{i_{\sigma(1)}j_{\sigma(1)}} M_{i_{\sigma(2)}j_{\sigma(2)}}}_N\cdots
\corr{M_{i_{\sigma(2n-1)} j_{\sigma(2n-1)}}M_{i_{\sigma(2n)}j_{\sigma(2n)}}}_N,
\een
where the summation on the right-hand
can be taken over the set of all possible ways of dividing
$\{1, \dots, 2n\}$ into $n$ pairs.
For example,
when $n=1$,
\ben
\corr{\frac{1}{2}\tr(M^2)}_N & = & \half \sum_{i,j=1}^N \corr{M_{ij}M_{ji}}_N
= \half \sum_{i,j=1}^N g_s \delta_{ii}\delta_{jj} = \half N^2 g_s,\\
\een
and when $n=2$,
\ben
\corr{\frac{1}{4}\tr(M^4)}_N &= & \frac{1}{4}
\sum_{i_1,\ldots, i_4=1}^N \corr{M_{i_1i_2}M_{i_2i_3}M_{i_3i_4}M_{i_4,i_1}}_N \\
& = & \frac{1}{4}\sum_{i_1,\ldots, i_4=1}^N \corr{M_{i_1i_2}M_{i_2i_3}}_N
\corr{M_{i_3i_4}M_{i_4,i_1}}_N \\
& + &\frac{1}{4}\sum_{i_1,\ldots, i_4=1}^N \corr{M_{i_1i_2}M_{i_3i_4}}_N
\corr{M_{i_2i_3}M_{i_4,i_1}}_N   \\
&+&\frac{1}{4}\sum_{i_1,\ldots, i_4=1}^N \corr{M_{i_1i_2}M_{i_4i_1}}_N
\corr{M_{i_2i_3}M_{i_3,i_4}}_N \\
& = & \frac{1}{4} \sum_{i_1, \ldots, i_4=1}^N g_s\delta_{i_1i_3} \delta_{i_2i_2} \cdot
g_s \delta_{i_3i_1}\delta_{i_4i_4} \\
& + &\frac{1}{4} \sum_{i_1,\ldots, i_4=1}^N g_s \delta_{i_1i_4}\delta_{i_2i_3} \cdot
g_s \delta_{i_2i_1}\delta_{i_3i_4} \\
& + & \frac{1}{4} \sum_{i_1, \ldots, i_4=1}^N g_s\delta_{i_1i_1} \delta_{i_2i_4} \cdot
g_s \delta_{i_2i_4}\delta_{i_3i_3} \\
& = & \frac{1}{2}g_s^2 N^3 +\frac{1}{4}g_s^2N.
\een
The contributions to the correlators can be represented by fat graphs.
For example,
$\tr(M^4)$ can be represented by
$$
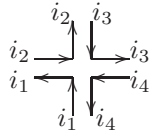
\xy
(-5,0); (0,0), **\dir{-};  ?>* \dir{>}; (0,5), **\dir{-};  ?>* \dir{>};
(2.5,5); (2.5,0), **\dir{-};  ?>* \dir{>};  (7.5,0), **\dir{-};  ?>* \dir{>};
(-7,1)*+{i_2}; (-7,-3)*+{i_1};  (-1.5,6)*+{i_2}; (4,6)*+{i_3};
(0,-7.5); (0,-2.5), **\dir{-};  ?>* \dir{>};  (-5,-2.5), **\dir{-};  ?>* \dir{>};
(7.5,-2.5); (2.5,-2.5), **\dir{-};  ?>* \dir{>};  (2.5,-7.5), **\dir{-};  ?>* \dir{>};
(9,1)*+{i_3}; (9,-3)*+{i_4};  (-0.6,-8)*+{i_1}; (4.5,-8)*+{i_4};
\endxy
$$
The two terms that contributes to $\corr{\frac{1}{4}\tr(M^4)}_N$ correspond to the
following two fat graphs:
The first is a fat graph with one $4$-valent vertex,
two edges, and three boundary components,
its automorphism group has order $2$,
and so its contribution is $\frac{1}{2}N^3g_s^2$:
$$
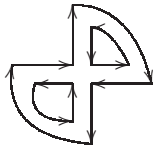
\xy
(-8,0); (0,0), **\dir{-};  ?>* \dir{>}; (0,8), **\dir{-};  ?>* \dir{>};
(2.5,5); (2.5,0), **\dir{-};  ?>* \dir{>};  (7.5,0), **\dir{-};  ?>* \dir{>};
(7.5,0); (2.5,5), **\crv{(5,5)}; ?>*\dir{>};
(0,8); (10.5,-2.5), **\crv{(8,8)}; ?>*\dir{>};
(0,-7.5); (0,-2.5), **\dir{-};  ?>* \dir{>};  (-5,-2.5), **\dir{-};  ?>* \dir{>};
(10.5,-2.5); (2.5,-2.5), **\dir{-};  ?>* \dir{>};  (2.5,-10.5), **\dir{-};  ?>* \dir{>};
(-5,-2.5); (0,-7.5), **\crv{(-6.5,-7.5)}; ?>*\dir{>};
(2.5,-10.5); (-8,0), **\crv{(-9.5,-9.5)}; ?>*\dir{>};
\endxy
$$
the second is a fat graph with one $4$-valent vertex,
two edges, and one boundary components,
its automorphism group has order $4$,
and so its contribution is $\frac{1}{4} Ng_s^2$:
$$
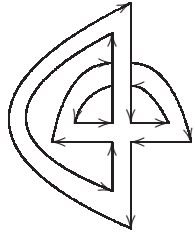
\xy
(-5,0); (0,0), **\dir{-};  ?>* \dir{>}; (0,12), **\dir{-};  ?>* \dir{>};
(2.5,16); (2.5,0), **\dir{-};  ?>* \dir{>};
(7.5,0), **\dir{-};  ?>* \dir{>};
(7.5,0); (2.5,5), **\crv{(5,5)}; ?>*\dir{>};
(0,5); (-5, 0), **\crv{(-4,4)}; ?>*\dir{>};
(-8,-2.5); (0,8), **\crv{(-6,8)}; ?>*\dir{>};
(2.5,8); (10.5,-2.5), **\crv{(8,8)}; ?>*\dir{>};
(0,-9); (0,-2.5), **\dir{-};  ?>* \dir{>};  (-8,-2.5), **\dir{-};  ?>* \dir{>};
(10.5,-2.5); (2.5,-2.5), **\dir{-};  ?>* \dir{>};  (2.5,-14), **\dir{-};  ?>* \dir{>};
(0,12); (0,-9),  **\crv{(-23,0)}; ?>*\dir{>};
(2.5,-14); (2.5,16),  **\crv{(-30,0)}; ?>*\dir{>};
\endxy
$$

Now suppose that $p_\lambda = p_{1}^{m_1} \cdots p_{k}^{m_k}$.
Then one has
\begin{align*}
l(\lambda) &= \sum_{i=1}^k m_i, &
|\lambda| & = \sum_{i=1}^k im_i, &
z_\lambda &= \prod_{i=1}^m i^{m_i}m_i!.
\end{align*}
Each $p_i$ is represented by a vertex where $i$ oriented bands
are glued together.
The contributions to
\ben
\corr{p_\lambda}= g_s^{-l(\lambda)}\corr{(\tr(M))^{m_1} \cdots (\tr(M^k))^{m_k}}_N
\een
is given by the following Fynman rules:
\be
\sum_{\Gamma^\lambda} \frac{w_\Gamma}{|\Aut(\Gamma)|},
\ee
where the summation is taken over the set $\Gamma^\lambda$  of all fat groups $\Gamma$
obtained by gluing the $l(\lambda)$ atoms specified by $\lambda$,
and for a fat graph $\Gamma$, $w_\Gamma$ takes the following form:
\be
w_\Gamma
= \prod_{v\in V(\Gamma)} w_v \cdot \prod_{e \in E(\Gamma)}w_e \cdot
\prod_{\in F(\Gamma)} w_f.
\ee
Here  $V(\Gamma)$ denotes the set of vertices of $\Gamma$,
and for $v\in V(\Gamma)$,
\be
w_v : = g_s^{-1},
\ee
$E(\Gamma)$ denotes the set of edges of $\Gamma$,
and for $e \in E(\Gamma)$,
\be
w_e : = g_s;
\ee
finally,
$F(\Gamma)$ denotes the set of boundary components of the surface underlying $\Gamma$,
and for $f \in \Gamma$,
\be
w_f : = N.
\ee
Therefore,
we actually have
\be
w_\Gamma = g_s^{|E(\Gamma)| - |V(\Gamma)|} N^{|F(\Gamma)|}.
\ee
For $\Gamma \in \Gamma^\lambda$,
it is clear that
\begin{align}
|V(\Gamma)| &= \sum_{i=1}^k m_i = l(\lambda), &
|E(\Gamma)| &= \half \sum_{i=1}^k im_i = \half |\lambda|,
\end{align}
and so  we have
\be
\corr{\frac{1}{z_\lambda} p_\lambda}_N
= \sum_{\Gamma \in \Gamma^\lambda}
\frac{1}{|\Aut(\Gamma)|} g_s^{\half|\lambda| - l(\lambda)} N^{|F(\Gamma)|}.
\ee

For example,
in degree two we have
\ben
&& \frac{1}{2}\corr{p_2}_N = \half N^2,\\
&& \frac{1}{2!}\corr{p_1^2}_N =\half Ng_s^{-1} ,
\een
in degree four,
\ben
&& \frac{1}{4} \corr{p_4}_N = (\frac{1}{2}N^3 + \frac{1}{2} N)g_s, \\
&& \frac{1}{3} \corr{p_3p_1}_N = N^2, \\
&& \frac{1}{2^22!}\corr{p_2^2}_N =  \frac{1}{4} N^2 + \frac{1}{8}N^4, \\
&& \frac{1}{2!2} \corr{p_2p_1^2}_N = (\frac{1}{2}N + \frac{1}{4}N^3)g_s^{-1} , \\
&& \frac{1}{4!} \corr{p_1^4}_N = \frac{1}{8} N^2g_s^{-2}.
\een

There is a similar formula for connected correlators $\corr{\frac{1}{z_\lambda}
p_\lambda}_N^c$:
\be \label{eqn:correlator}
\corr{\frac{1}{z_\lambda} p_\lambda}_N^c
= \sum_{\Gamma \in \Gamma^{\lambda c}}
\frac{1}{|\Aut(\Gamma)|} g_s^{\half|\lambda| - l(\lambda)} N^{|F(\Gamma)|},
\ee
where $\Gamma^{\lambda c}$ is the set of connected fat graphs of type $\lambda$.
From this it is easy to see that
$\corr{p_\lambda}_N^c|_{g_s=1}$ is a polynomial in $N$ with nonnegative integer as
coefficients,
this is because for $\Gamma\in \Gamma^{\lambda, c}$,
\be
\frac{z_\lambda}{|\Aut(\Gamma)}\in \bZ.
\ee

\subsection{Calculations of the correlators by representation theory of symmetric groups}
\label{sec:Itz-Zub}

For higher degrees,
it becomes very complicated to write down all the possible fat graphs and
compute the orders of their automorphism groups.
So it is desirable to find other methods to evaluate $\corr{p_\lambda}_N$.
We now recall the calculations of $\corr{p_\lambda}_N$ by
representation theory of the symmetric group due to Itzykson and Zuber \cite{Itz-Zub}.

First of all,
given a fat graph $\Gamma$,
one can consider the (thin) graph $\hat{\Gamma}$ obtained from it by retracting each
band to a one-dimensional edge.
To get a fat graph $\Gamma$ from a thin graph $\hat{\Gamma}$,
it suffices to fix a cyclic ordering of the edges incident at vertex.
Therefore,
given a fat graph $\Gamma \in \Gamma^\lambda$,
where $|\lambda| = 2n$,
cut each edge of $\hat{\Gamma}$ in half,
label the $2n$ half edges by  $1, \ldots, 2n$.
Using the cyclic ordering at each vertex,
each such ordering determines a permutation $\sigma$ in $S_{2n}$
in the conjugacy class $C_\lambda$.
Each edge determines a transposition $\tau_e= (i_e,j_e)$,
where $i_e, j_e$ are the labellings of two half-edges obtained by cutting $e$.
The involutions $\{\tau_e\}_{e\in E(\Gamma)}$ are disjoint,
let $\tau = \prod_{e\in E(\Gamma)} \tau_e$.
It is of type $C_{(2^n)}$.
Now consider the labelling of the fat graph $\Gamma$
induced from that of $\hat{\Gamma}$ by assigning the
same number on each side of the half-band.
By following the arrows along the boundary components
of $\Gamma$,
one gets a partition of $2n$ of length $|F(\Gamma)|$.
From such considerations,
Itzykson and Zuber \cite[(2.7)]{Itz-Zub} obtained:
\be \label{eqn:IZ}
\corr{p_\lambda}_N = \frac{1}{|C_\lambda|} g_s^{|\lambda|/2-l(\lambda)} \sum_{\mu \in \cP_{2n}}
\sum_{\tau \in C_{(2^n)}, \sigma \tau \in C_\mu} N^{l(\mu)}.
\ee
So one needs to count the number $N_{C_\lambda, C_{(2^n)}}^{C_\mu}$ of solutions of the following equation:
\be
\sigma \tau \in C_\mu, \;\; \sigma \in C_\lambda, \;\; \tau \in C_{(2^n)}.
\ee
Now we recall how to solve such an equation using representation theory.

For a finite group $G$, given three conjugacy classes $C_1, C_2, C_3$,
consider the number $N_{C_1,C_2, C_3}$ of solutions
\be
g_1g_2 g_3 = e, \;\; g_i \in C_i, \; i=1, 2,3.
\ee
Write $\sigma_i = \sum_{g\in C_i} g$.
This is an element in the center of the group ring $\bC G$.
Hence by Schur's lemma, it acts a constant on any irreducible representation $V^\rho$
of $G$,
and since
\be
\tr \sigma_i|_{V^\rho} = |C_i| \cdot \chi^\rho|_{C_i},
\ee
where $\chi^\rho$ is the character of $V^\rho$, therefore,
\be
\sigma_i|_{V^\rho} = \frac{|C_i|}{\dim V^\rho} \chi^\rho|_{C_i},
\ee
and so
\be \label{eqn:123}
\sigma_1\sigma_2\sigma_3|_{V^\rho} = \prod_{i=1}^3 \frac{|C_i|}{\dim V^\rho} \chi^\rho|_{C_i}.
\ee
Now consider the action of $\sigma_1\sigma_2\sigma_3$ on
the regular representation $\bC G$ of $G$.
On the one-hand,
because
\be
\tr g|_{\bC G} = |G| \cdot \delta_{g,e},
\ee
we have
\be
\tr \sigma_1\sigma_2\sigma_3|_{|bC G} = |G| \cdot N(C_1, C_2, C_3);
\ee
on the other hand,
using the decomposition
\be
\bC G = \bigoplus_{\rho \in G^\vee} V^\rho \otimes \bC^{\dim V^\rho},
\ee
where $G^\vee$ is the set of equivalence classes of irreducible representations of $G$,
by \eqref{eqn:123} one gets:
\ben
\tr \sigma_1\sigma_2\sigma_3|_{\bC G}
& = & \sum_{\rho \in G^\vee} ( \dim V^\rho) \cdot \tr \sigma_1\sigma_2\sigma_3|_{V^\rho} \\
& = &  \sum_{\rho \in G^\vee} \frac{1}{\dim V^\rho} \prod_{i=1}^3 |C_i| \cdot \chi^\rho|_{C_i}.
\een
Therefore,
\be
N_{C_1, C_2, C_3} = \frac{1}{|G|} \sum_{\rho  \in G^\vee} \frac{1}{\dim V^\rho} \prod_{i=1}^3 |C_i| \cdot \chi^\rho|_{C_i}.
\ee
This is a special case of the Burnside formula.

Let us recall some well-known facts from the representation theory
of the symmetric groups to fix the notations.
The conjugacy classes of the symmetric group $S_n$
are in one-to-one correspondence with partitions $\lambda \in \cP_n$
(the set of partitions of $n$).
Denote by $C_\lambda$ the number of elements in the class with cycle type $\lambda$.
Each element in this class consists of $l$ disjoint cycles, with lengths
$\lambda_1, \dots, \lambda_l$ respectively.
The number of elements in this class is
\be
\frac{n!}{z_\lambda} = \frac{n!}{1^{m_1}m_1! \cdots n^{m_n} m_n!}.
\ee
The irreducible characters  of $S_n$
are also indexed by partitions $\lambda\in \cP_n$.
Denote them by $\chi^\lambda$.
The value of $\chi^\lambda$ on the conjugacy class $C_\mu$ is denoted by $\chi^\lambda_\mu$.
They are all real numbers and they satisfy
the orthogonality relations:
\bea
&& \sum_{\lambda\in \cP_n}  \chi^\lambda_\mu \chi^\lambda_\nu
= \frac{n!}{z_\lambda} \delta_{\mu, \nu}, \\
&& \frac{1}{n!} \sum_{\tau \in S_n} \chi^\lambda(\tau) \chi^\mu(\sigma\tau)
= \frac{\chi^\lambda(\sigma)}{\chi^\lambda_{(1^n)}} \delta_{\lambda, \mu}.
\eea

It is easy to see that $C_\mu^{-1} = C_\mu$.
By Burnside formula,
one then gets \cite[(2.7)]{Itz-Zub}:
\be
\frac{1}{z_\lambda} \corr{p_\lambda}_N
= g_s^{|\lambda|/2-l(\lambda)} \sum_{\mu \in \cP_{2n}} N^{l(\mu)} \sum_{\nu \in \cP_{2n}}
\frac{(2n)!}{z_\lambda z_{(2^n)}z_{\mu}}
\frac{\chi^\nu_{(2^n)}\chi^\nu_{\mu} \chi^\nu_{\lambda}}{\chi^\nu_{(1^{2n})}}.
\ee
With this formula,
one can compute more correlators.
For example,
  in degree six:
\ben
&& \corr{p_6}_N = (10 N^2 + 5 N^4)g_s^2, \\
&& \corr{p_5p_1}_N = (5N + 10N^3)g_s, \\
&& \corr{p_4p_2}_N = (4N + 9 N^3 + 2N^5) g_s, \\
&& \corr{p_4p_1^2}_N = 13N^2 + 2N^4, \\
&& \corr{p_3^2}_N = (3N + 12N^3)g_s, \\
&& \corr{p_3p_2p_1}_N = 12N^2 + 3N^4, \\
&& \corr{p_3p_1^3}_N = (6N+9N^3) g_s^{-1}, \\
&& \corr{p_2^3}_N = 8N^2+6N^4+N^6, \\
&& \corr{p_2^2p_1^2}_N = (8N+6N^3+ N^5)g_s^{-1}, \\
&& \corr{p_2p_1^4}_N = (12N^2 + 3N^4) g_s^{-2}, \\
&& \corr{p_1^6}_N = 15 N^3g_s^{-3}.
\een
Combining with the examples in degree two and degree four in \S \ref{sec:Feynman},
the first few terms of the partition function are:
\ben
Z_N & = & 1 + N^2\cdot \frac{g_2}{2} +  Ng_s^{-1} \cdot \frac{g_1^2}{2}
+  (N + 2 N^3)g_s \cdot \frac{g_4}{4} + 3N^2 \cdot \frac{g_3g_1}{3} \\
& + & (2N^2 + N^4) \cdot \frac{g_2^2}{8} + (2N + N^3)g_s^{-1} \cdot \frac{g_2g_1^2}{4}
+ 3N^2 g_s^{-2} \cdot \frac{g_1^4}{4!} \\
& + & (10 N^2 + 5 N^4) g_s^2 \frac{g_6}{6}
+ (5N + 10N^3)g_s \cdot \frac{g_5g_1}{5} \\
& + & (4N + 9 N^3 + 2N^5)g_s \cdot \frac{g_4g_2}{8}
+ (13N^2 + 2N^4) \cdot \frac{g_4g_1^2}{8} \\
& + & (3N + 12N^3)g_s \cdot \frac{g_3^2}{18}
+ (12N^2 + 3N^4) \cdot \frac{g_3g_2g_1}{6} \\
& + & (6N+9N^3)g_s^{-1} \cdot \frac{g_3g_1^3}{18}
+ (8N^2+6N^4+N^6) \cdot \frac{g_2^3}{48} \\
& + & (8N+6N^3+ N^5) g_s^{-1} \cdot \frac{g_2^2g_1^2}{16}
+ (12N^2 + 3N^4) g_s^{-2} \cdot \frac{g_2g_1^4}{48} \\
& + & 15 N^3 g_s^{-3} \cdot \frac{g_1^6}{720} + \cdots.
\een
Note
\ben
Z_N &= & \sum_\lambda \frac{1}{z_\lambda}\corr{p_\lambda}_N \prod_i g_{\lambda_i} \\
& = & \sum_\lambda  g_s^{|\lambda|/2-l(\lambda)}
\sum_{\mu \in \cP_{2n}} N^{l(\mu)} \sum_{\nu \in \cP_{2n}}
\frac{(2n)!}{z_\lambda z_{(2^n)}z_{\mu}}
\frac{\chi^\nu_{(2^n)}\chi^\nu_{\mu}
\chi^\nu_{\lambda}}{\chi^\nu_{(1^{2n})}} \prod_i g_{\lambda_i} \\
& = & \sum_\lambda    \sum_{\mu \in \cP_{2n}} N^{l(\mu)} \sum_{\nu \in \cP_{2n}}
\frac{(2n)!}{z_\lambda z_{(2^n)}z_{\mu}}
\frac{\chi^\nu_{(2^n)}\chi^\nu_{\mu} \chi^\nu_{\lambda}}{\chi^\nu_{(1^{2n})}}
\prod_i (\lambda^{\lambda_i/2-1}g_{\lambda_i}).
\een
So no information will be lost if one takes $g_s =1$,
because one can recover $Z_N$ from $Z_N|_{g_s=1}$ by simply
changing $p_n$ to $\lambda^{n/2-1}p_n$.

\section{Closed Formula for $n$-Point Functions of Hermitian Matrix Model}
\label{sec:Sato}

In this section we use the fact that $Z_N$ is a $\tau$-function
of the KP hierarchy
and apply the formula in \cite{Zhou}
for $n$-point functions associated with a $\tau$-function of the KP hierarchy.

\subsection{The $n$-point functions of Hermitian matrix model}
After taking the logarithm of  $Z_N$,
one gets the first few terms of the free energy $F_N$ as follows:
\ben
F_N & = & \frac{1}{2}N^2g_2+\frac{1}{2}Ng_s^{-1}g_1^2
+(\frac{1}{2}N^3+\frac{1}{4}N)g_s g_4+N^2g_3g_1+\frac{N^2}{4}g_2^2
+ \frac{N}{2}g_s^{-1} g_2g_1^2 \\
& + & (\frac{5N^2}{3}+\frac{5N^4}{6})g_s^2g_6
 +(N+2N^3)g_sg_5g_1 + (\frac{N}{2}+N^3) g_s g_4g_2 + \frac{3N^2}{2} g_4g_1^2 \\
& + & \biggl(\frac{N}{6} +\frac{2N^3}{3} \biggr) g_s g_3^2
+  2N^2 g_3g_2g_1 + \frac{N}{3}g_s^{-1}g_3g_1^3
+\frac{N^2}{6} g_2^3 + \frac{N}{2} g_s^{-1} g_2^2g_1^2 + \cdots
\een
From this one sees that
the first few terms of the $n$-point part of the free energy  for $n=1, 2,3$ are
\ben
F_N^{(1)}  & = & \frac{1}{2}N^2g_2
+(\frac{1}{2}N^3+\frac{1}{4}N)g_s g_4
+ (\frac{5N^2}{3}+\frac{5N^4}{6})g_s^2g_6 + \cdots, \\
F_N^{(2)}  & = & \frac{1}{2}Ng_s^{-1}g_1^2
+N^2g_3g_1+\frac{N^2}{4}g_2^2
 +(N+2N^3)g_sg_5g_1 \\
& + &  (\frac{N}{2}+N^3) g_s g_4g_2
+ \biggl(\frac{N}{6} +\frac{2N^3}{3} \biggr) g_s g_3^2  + \cdots, \\
F_N^{(3)}  & = &   \frac{N}{2}g_s^{-1} g_2g_1^2
+  \frac{3N^2}{2} g_4g_1^2
+  2N^2 g_3g_2g_1
+\frac{N^2}{6} g_2^3   + \cdots.
\een

Since we are only interested in the coefficients of $F^{(n)}_N$,
we encode them in the $n$-point function defined by:
\be
G_N^{(n)}(\xi_1, \ldots, \xi_n):
= \sum_{j_1, \dots, j_n \geq 1} \frac{\pd^n F^{(n)}}{\pd T_{j_1}
\cdots \pd T_{j_n}}\biggl|_{g_s=1} \cdot \xi_1^{-j_1-1}\cdots \xi_n^{-j_n-1},
\ee
where
\be
T_n = \frac{g_n}{n}.
\ee
For example,
\ben
G_N^{(1)}(\xi_1) & = & N^2\xi_1^{-3}
+(2N^3+N)\xi_1^{-5}
+ (10N^2+5N^4)\xi_1^{-7} + \cdots, \\
G_N^{(2)}(\xi_1,\xi_2) & = & N\xi_1^{-2}\xi_2^{-2}
+ 3N^2(\xi_1^{-2}\xi_2^{-4} + \xi_1^{-4} \xi_2^{-2})
+ 2N^2\xi_1^{-3}\xi_2^{-3} \\
& + & 5 (N+2N^3) (\xi_1^{-2} \xi_2^{-6} + \xi_1^{-6}\xi_2^{-2})  \\
& + &  (4N+8N^3) (\xi_1^{-3} \xi_2^{-5} + \xi_1^{-5}\xi_2^{-3})
+ \biggl(3N + 12N^3 \biggr) \xi_1^{-4}\xi_2^{-4} + \cdots, \\
G_N^{(3)}(\xi_1,\xi_2,\xi_3) & = &
2 N (\xi_1^{-2}\xi_2^{-2} \xi_3^{-3}+\xi_1^{-2}\xi_2^{-3}\xi_3^{-2}
+\xi_1^{-3}\xi_2^{-2}\xi_3^{-2}) \\
& + & 12 N^2 (\xi_1^{-2}\xi_2^{-2} \xi_3^{-5}
+\xi_1^{-2}\xi_2^{-5}\xi_3^{-2}+\xi_1^{-5}\xi_2^{-2}\xi_3^{-2}) \\
& + & 12N^2 (\xi_1^{-2}\xi_2^{-3} \xi_3^{-4}+\xi_1^{-2}\xi_2^{-4}\xi_3^{-3}
+\xi_1^{-3}\xi_2^{-2}\xi_3^{-4} \\
&&+ \xi_1^{-3}\xi_2^{-4} \xi_3^{-2}+\xi_1^{-4}\xi_2^{-2}\xi_3^{-3}
+\xi_1^{-4}\xi_2^{-3}\xi_3^{-2}) \\
& + & 8 N^2 \xi_1^{-3}\xi_2^{-3} \xi_3^{-3}    + \cdots.
\een

One can recover $F^{(n)}_N$ from $G^{(n)}_N$ as follows:
\be
F^{(n)}_N = \frac{1}{n!} G^{(n)}_N(\xi_1, \ldots, \xi_n)|_{\xi_i^{-j-1} \mapsto g_j/j}.
\ee
Therefore,
giving a closed formula for $G^{(n)}_N$ is equivalent to
giving a closed formula for $F^{(n)}_N$.

It is not practical to compute the $n$-point functions of the Hermitian
one-matrix model by either counting the fat graphs or computing the characters
of all symmetric groups.
Other methods have been developed for this purpose.
See e.g. \cite{ACKM, Eyn-Ora,Har-Zag, Mor-Sha}.
More recently,
Dubrovin and Yang \cite{Dub-Yan} have proved a formula for $n$-point functions
of Hermitian one-matrix models based on the connection with Toda lattice hierarchy.
We will present below a different formula based on connection with
the KP hierarchy.

\subsection{Partition function of Hermitian matrix model as $\tau$-function of KP hierarchy}

It is well-known that the partition functions $\{Z_N\}_{n \geq 1}$ give
a tau-function of the Toda lattice hierarchy of Ueno and Takasaki \cite{UT}.
See e.g. Gerasimov {\em et al} \cite{GMMMO}.
This is the starting point for the computations of
Dubrovin-Yang \cite{Dub-Yan}.
By a result of Shaw-Tu-Yen \cite{Shaw-Tu-Yen},
$Z_N$ is a tau-function of the KP hierarchy
with respect to $T_1, T_2, \ldots$, where $T_n = g_n/n$.
(See also Mulase \cite{Mulase}.)
This fact is the point of departure for our computation for $n$-point functions
of Hermitian matrix model.

\subsection{Formula for $n$-point function associated with $\tau$-function
of KP hierarchy}
\label{sec:Formula}

In an early work \cite{Zhou},
we have obtained a formula for computing $n$-point
functions associated with any tau-function (in formal power series)
of the KP hierarchy.
Let us recall this formula.
Suppose that the tau-function corresponds to an element in Sata Grasssmannian
$U \in Gr^{(0)}$, given by a normalized basis
$$\{f_n = z^{n} + \sum_{m \geq 0} a_{n,m} z^{-m - 1} \}_{n \geq 0},$$
then after the boson-fermion correspondence the tau-function corresponds to:
\ben
|U\rangle = e^A \vac,
\een
where $A$ is a linear operator on the fermionic Fock space:
\ben
A = \sum_{m, n \geq 0} a_{n,m} \psi_{-m-1/2} \psi^*_{-n-1/2}.
\een
Furthermore,
for $n \geq 1$,
the bosonic $n$-function associated to tau-function of the KP hierarchy is:
\ben
&& \sum_{j_1,\dots, j_n \geq 1}
\frac{\pd^n F_U}{\pd T_{j_1} \cdots \pd T_{j_n} } \biggl|_{\bT =0}
  \xi_1^{-j_1-1}\cdots \xi_n^{-j_n-1} + \frac{\delta_{n,2}}{(\xi_1-\xi_2)^2} \\
& = & (-1)^{n-1} \sum_{\text{$n$-cycles}}  \prod_{i=1}^n \hat{A}(\xi_{\sigma(i)}, \xi_{\sigma(i+1)}),
\een
where $\hat{A}(\xi_i, \xi_j)$ are the propagators:
\ben
\hat{A}(\xi_i, \xi_j) = \begin{cases}
i_{\xi_i, \xi_j} \frac{1}{\xi_i-\xi_j} + A(\xi_i, \xi_j),  & i < j, \\
A(\xi_i, \xi_i),  & i =j, \\
i_{\xi_j, \xi_i} \frac{1}{\xi_i-\xi_j} + A(\xi_i, \xi_j),  & i > j.
\end{cases}
\een
In the above we have used the following notations:
\ben
&& A(\xi, \eta) =  \sum_{m,n\geq 0} a_{n,m} \xi^{-m-1} \eta^{-n-1}, \\
&& i_{x, y} \frac{1}{(x-y)^n} = \sum_{k \geq 0} \binom{-n}{k} x^{-n-k} y^k.
\een

\subsection{Partition function $Z_N$ in representation basis}
\label{sec:Gro-Tay}

In order to apply the formula in last subsection,
we need to rewrite $Z_N$ in terms of Schur functions.
We will first consider $Z_N|_{g_s=1}$,
and regard $g_n$ as the Newton power functions $p_n$.
Recall
Schur functions $\{s_\lambda\}$ and the Newton functions $\{p_\mu \}$ are related to each other
by the Frobenius formula:
\bea
&& p_\mu = \sum_\lambda \chi^\lambda_\mu s_\lambda, \label{eqn:Newton-in-Schur} \\
&& s_\lambda = \sum_\mu \frac{\chi^\lambda_\mu}{z_\mu} p_\mu. \label{eqn:Schur-in-Newton}
\eea
Then we have
\be
Z_N|_{g_s=1} = \sum_\lambda \corr{s_\lambda}_N\cdot s_\lambda.
\ee

By \eqref{eqn:IZ}, we can derive the following formula:
\be
\corr{s_\lambda}_N =(2n-1)!! \frac{\chi^\lambda_{(2^n)}}{\chi^\lambda_{(1^{2n})}} \cdot
\sum_{\mu \in \cP_{2n}} \chi^\lambda_\mu \frac{N^{l(\mu)}}{z_\mu}.
\ee
Indeed,
\ben
\corr{s_\lambda}_N
& = & \sum_{\eta\in \cP_{2n}} \frac{\chi^\lambda_\eta}{z_\eta} \corr{p_\eta}_N \\
& = & \sum_{\eta\in \cP_{2n}} \frac{\chi^\lambda_\eta}{z_\eta}
\sum_{\mu \in \cP_{2n}} N^{l(\mu)} \sum_{\nu \in \cP_{2n}}
\frac{(2n)!}{z_{(2^n)}z_{\mu}}
\frac{\chi^\nu_{(2^n)}\chi^\nu_{\mu} \chi^\nu_{\eta}}{\chi^\nu_{(1^{2n})}} \\
& = &
\sum_{\mu \in \cP_{2n}} N^{l(\mu)} \sum_{\nu \in \cP_{2n}}
\frac{(2n)!}{2^nn!z_{\mu}}
\frac{\chi^\nu_{(2^n)}\chi^\nu_{\mu} }{\chi^\nu_{(1^{2n})}} \delta_{\lambda,\nu} \\
& = & (2n-1)!! \frac{\chi^\lambda_{(2^n)}}{\chi^\lambda_{(1^{2n})}} \cdot
\sum_{\mu \in \cP_{2n}} \chi^\lambda_\mu \frac{N^{l(\mu)}}{z_\mu}.
\een

By \cite[Example 4, p. 45]{Macdonald},
\ben
\sum_{\mu \in \cP_{2n}} \chi^\lambda_\mu \frac{N^{l(\mu)}}{z_\mu}
= s_\lambda|_{p_n=N}
= \prod_{x\in \lambda} \frac{N+c(x)}{h(x)},
\een
where $c(x)$ and $h(x)$ denotes the content and the hook length of $x$
respectively.
For $N$ large enough,
the right-hand side is the dimension of the irreducible representation of $U(N)$
indexed by $\lambda$ (cf. \cite{Marino}).
In fact,
the dimension of a representation of $U(N)$
whose Young tableaux has rows of length
$(\lambda_1, \lambda_2, \dots, \lambda_l)$ is given by Weyl's formula,
\be
\dim R_\lambda^{U(N)} =
\frac{\prod_{1\leq i<j \leq N} (h_i - h_j)}{\prod_{1\leq i<j \leq N} (j - i)},
\ee
where $h_i = N + \lambda_i - i$.
One can separate the product into three cases:
\ben
\dim R_\lambda^{U(N)} & = &
\frac{\prod_{1\leq i<j \leq l} (h_i - h_j)}{\prod_{1\leq i<j \leq l} (j - i)}
\cdot \frac{\prod_{1\leq i\leq l <j \leq N } (h_i - h_j)}{\prod_{1\leq i\leq l <j \leq N} (j - i)}
\cdot \frac{\prod_{l< i<j \leq N} (h_i - h_j)}{\prod_{l < i<j \leq N} (j - i)} \\
& = & \prod_{1 \leq i < j \leq l} \frac{(\lambda_i-i) - (\lambda_j-j)}{j-i}
\cdot \prod_{1 \leq i \leq l < j \leq N} \frac{(\lambda_i-i) + j}{j-i} \\
& = & \prod_{1 \leq i < j \leq l} \frac{(\lambda_i-i) - (\lambda_j-j)}{j-i}
\cdot \prod_{1 \leq i \leq l} \prod_{j=l+1}^N\frac{ (\lambda_i-i) + j}{j-i} \\
& = & \prod_{1 \leq i < j \leq l} \frac{(\lambda_i-i) - (\lambda_j-j)}{j-i}
\cdot \prod_{1 \leq i \leq l} \frac{ (\lambda_i-i + N)!(l-i)!}{(\lambda_i-i + l)!(N-i)!} \\
& = & \prod_{1 \leq i < j \leq l} \frac{(\lambda_i-i) - (\lambda_j-j)}{j-i}
\cdot \prod_{1 \leq i \leq l} \frac{(l-i)!}{(\lambda_i-i + l)!}
\cdot \prod_{1 \leq i \leq l} \prod_{j=1}^{\lambda_i} (N+j-i) \\
& = & \frac{\prod_{1 \leq i < j \leq l} (\lambda_i-i) - (\lambda_j-j)}{
\prod_{1 \leq i \leq l}  (\lambda_i-i + l)!}
\cdot \prod_{x\in \lambda}  (N+c(x)) \\
& = & \prod_{x\in \lambda}  \frac{(N+c(x))}{h(x)}.
\een
Here in the last equality we have used Macdonald
\cite[p. 11, (4)]{Macdonald}.

Let $R_\lambda$ be the irreducible representation of
 the symmetric group of $|\lambda|:=\sum_{i=1}^l \lambda_i$ objects,
corresponding to the partition $\lambda$.
Then one has
\be
d_\lambda:= \dim R_\lambda = \chi^\lambda_{(1^{|\lambda|})}
= \frac{\prod_{v\in \lambda} h(v)}{|\lambda|!}.
\ee
As observed by Gross \cite[Appendix A1]{Gross}
(see also Gross-Taylor \cite[Appendix A]{Gross-Taylor}),
\be
\dim R_\lambda^{U(N)} = \frac{d_\lambda N^{|\lambda|}}{|\lambda|!}
\prod_{v\in \lambda}
\biggl(1 + \frac{c(v)}{N}  \biggr),
\ee
where $c(v)$ is the content of the box $v$ in the Young diagram $\lambda$:
If $v$ is at the $i$-th row and the $j$-column,
then $c(v) = j - i$.
Therefore,
\be
\corr{s_\lambda}_N = \frac{(2n-1)!!}{(2n)!} \chi^\lambda_{(2^n)}
\prod_{v \in \lambda} (N + c(v)).
\ee
Alternatively,
\be
\corr{s_\lambda }_{N} = c(\lambda) \dim R^{U(N)}_\lambda
= c(\lambda) \cdot \frac{d_\lambda N^{|\lambda|}}{|\lambda|!}
\prod_{v\in \lambda}
\biggl(1 + \frac{c(v)}{N}  \biggr),
\ee
where $c(\lambda)$ is defined by:
\be
c(\lambda):=(2n-1)!! \frac{\chi^\lambda_{(2^n)}}{\chi^\lambda_{(1^{2n})}}.
\ee
It is interesting to note that the dimension formula
for $\dim R_\lambda^{U(N)}$ and $\dim R_\lambda$ appear in
both matrix model theory and large $N$ Yang-Mills theory.

The explicit formula for the numbers $c(\lambda)$
was due to Di Francesco and Itzykson \cite{DiF-Itz}.
Define the set of $2n$
integers $f_i$ as follows
\be
f_i = \lambda_i + 2n - i, i = 1,\dots, 2n.
\ee
Following  these authors, we will say that $\lambda$ is even if
the number of odd $f_i$'s is the same as the number of even $f_i$'s.
Otherwise, we will say that it is odd.
One has the following result \cite{DiF-Itz}:
\be
c(\lambda) = \begin{cases}
 (-1)^{n(n-1)/2}
\frac{\prod_{f \;\; odd} f!! \prod_{f' \;\; even} (f'-1)!!}
{\prod_{f \;\; odd,f' \;\; even}(f - f')}, &\text{if $\lambda$ is even}, \\
0, &\text{otherwise}.
\end{cases}
\ee
We conjecture that any partition of an even number
is even in the sense of   \cite{DiF-Itz}.

For example,
\ben
&& \corr{s_{(2)}}_N = \half N(N+1), \\
&& \corr{s_{(1^2)}}_N = -\half N(N -1), \\
&& \corr{s_{(4)}}_N = \frac{1}{8}N(N+1)(N+2)(N+3), \\
&& \corr{s_{(3,1)}}_N = - \frac{1}{8}N(N+1)(N+2)(N-1), \\
&& \corr{s_{(2,2)}}_N = \frac{1}{4} N^2(N+1)(N-1), \\
&& \corr{s_{(2,1^2)}}_N = -\frac{1}{8} N(N+1)(N-1)(N-2), \\
&& \corr{s_{(1^4)}}_N = \frac{1}{8} N(N-1)(N-2)(N-3), \\
&& \corr{s_{(6)}}_N = \frac{1}{48}N(N+1)(N+2)(N+3)(N+4)(N+5), \\
&& \corr{s_{(5,1)}}_N = - \frac{1}{48} N(N+1)(N+2)(N+3)(N+4)(N-1), \\
&& \corr{s_{(4,2)}}_N = \frac{1}{16} N(N+1)(N+2)(N+3)(N-1)N, \\
&& \corr{s_{(4,1^2)}}_N = -\frac{1}{24} N(N+1)(N+2)(N+3)(N-1)(N-2), \\
&& \corr{s_{(3,3)}}_N = - \frac{1}{16} N(N+1)(N+2)(N-1)N(N+1), \\
&& \corr{s_{(3,2,1)}}_N = 0, \\
&& \corr{s_{(3,1,1,1)}}_N = \frac{1}{24} N(N+1)(N+2)(N-1)(N-2)(N-3), \\
&& \corr{s_{(2,2,2)}}_N = \frac{1}{16} N(N+1)(N-1)N(N-2)(N-1), \\
&& \corr{s_{2,2,1,1)}}_N = - \frac{1}{16} N(N+1)(N-1)N(N-2)(N-3), \\
&& \corr{s_{(2,1,1,1,1)}}_N = \frac{1}{48} N(N+1)(N-1)(N-2)(N-3)(N-4), \\
&& \corr{s_{(1,1,1,1,1,1)}}_N = - \frac{1}{48}N(N-1)(N-2)(N-3)(N-4)(N-5).
\een

To summarize,
\be
Z_N|_{g_s=1}
= \sum_\lambda c(\lambda) \cdot \prod_{x\in \lambda}  \frac{(N+c(x))}{h(x)} \cdot s_\lambda.
\ee

\subsection{The partition function $Z_N$ as a Bogoliubov transform}

By the general theory developed in \cite{Zhou},
we have
\be
Z_N|_{g_s=1} = e^A \vac
\ee
in fermionic picture,
where
\be
A = \sum_{m, n \geq 0}A_{m,n}\psi_{-m-1/2} \psi_{-n-1/2}^*.
\ee
In particular,
$Z_N|_{g_s=1}$ contains terms of the form
$A_{m,n}\psi_{-m-1/2} \psi_{-n-1/2}^*\vac$,
which after boson-fermion correspondence
corresponds to
\be
(-1)^n A_{m,n} s_{(m|n)},
\ee
where $(m|n)$ is a hook diagram in Frobenius notation:
it is the partition $(m+1, 1^n)$.
So one has
\be
(-1)^n A_{m,n} = \corr{s_{(m+1, 1^n)}}_N.
\ee
The right-hand has been computed by Itzykson and Zuber \cite{Itz-Zub}.
When $\lambda$ is a hook diagram of type $(p, q)$, i.e.,
$\lambda = (q+1, 1^p)$, such that $p+1+q = 2n$,
\be
\chi^{(q+1,p)}_{(2^n)} = (-1)^{[(p+1)/2]} \binom{n-1}{[p/2]},
\ee
where $[x]$ denotes the integral part of $x$,
and one has
\be
\corr{s_{(q+1,1^p)}}_N = (-1)^{[(p+1)/2]} \binom{n-1}{[p/2]}
\frac{(2n-1)!!}{(2n)!} [N]_{-p}^q .
\ee
Here, to simplify the notations,
we define for $k\leq l$,
\be
[N]_k^l:=\prod_{j=k}^l (N+j).
\ee
The basic properties of $[N]_k^l$ are
\bea
&& [N-1]_k^l = [N]_{k-1}^{l-1},\\
&& [N]_k^l-[N]_{k-1}^{l-1} = (l-k+1)\cdot [N]_k^{l-1}
= (l-k+1)\cdot [N-1]_{k+1}^l.
\eea

Therefore,
we get the following result:

\begin{thm} \label{thm:Main}
The fermionic representation of $Z_N$ is given explicitly as follows:
\be
\begin{split}
Z_N|_{g_s=1} = & \exp \biggl( \sum_{n=1}^\infty \frac{(2n-1)!!}{(2n)!} \sum_{p=0}^{2n-1}
(-1)^p \cdot (-1)^{[(p+1)/2]} \binom{n-1}{[p/2]} \\
& \cdot [N]_{-p}^{2n-1-p}  \cdot \psi_{-(2n-p)-1/2} \psi_{-p-1/2}^*
 \biggr) \vac.
\end{split}
\ee
\end{thm}

\subsection{The $n$-point function for Hermitian matrix model}

In last subsection we have  seen that
\be
Z_N = \exp (\sum_{p,q\geq 0} A_{q,p}\psi_{-q-1/2}\psi^*_{-p-1/2})\vac,
\ee
where the coefficients $A_{q,p}$ are explicitly given by:
\be
A_{q,p}
= \begin{cases}
(-1)^{p+[(p+1)/2]} \frac{(2n-1)!!}{(2n)!}
 \binom{n-1}{[p/2]} \cdot [N]_{-p}^{2n-1-p}, &
p+q=2n-1, \\
0, & \text{otherwise}.
\end{cases}
\ee
Hence one can apply our formula for $n$-point functions associated with
tau-functions of KP hierarchy.

\begin{thm} \label{thm:Main2}
The $n$-point function associated with $Z_N$ is given by the following formula:
\be
G_N^{(n)}(\xi_1, \ldots, \xi_n)
= (-1)^{n-1} \sum_{\text{$n$-cycles}}
\prod_{i=1}^n \hat{A}(\xi_{\sigma(i)}, \xi_{\sigma(i+1)})
-  \frac{\delta_{n,2}}{(\xi_1-\xi_2)^2},
\ee
where
\be \label{eqn:A}
\begin{split}
A(\xi, \eta)
= & \sum_{n \geq 1} \frac{(2n-1)!!}{(2n)!} \sum_{p=0}^{2n-1}
(-1)^{p+[(p+1)/2]} \binom{n-1}{[p/2]} \\
& \cdot [N]_{-p}^{2n-1-p} \cdot \xi^{-p-1}
\eta^{-(2n-1-p)-1}.
\end{split}
\ee
\end{thm}

The following are the first few terms of $A(\xi, \eta)$:
\ben
A(\xi, \eta)
& = & \half [N]_0^1\cdot \xi^{-1}\eta^{-2}
+ \half [N]_{-1}^0 \cdot \xi^{-2}\eta^{-1}
+ \frac{1}{8}[N]_0^3\cdot \xi^{-1} \eta^{-4} \\
& + & \frac{1}{8}[N]_{-1}^2 \cdot \xi^{-2} \eta^{-3}
-  \frac{1}{8} [N]_{-2}^1 \cdot \xi^{-3} \eta^{-2}
-  \frac{1}{8} [N]_{-3}^0 \cdot \xi^{-4} \eta^{-1} \\
& + & \frac{1}{48}[N]_0^5 \cdot \xi^{-1} \eta^{-6}
+ \frac{1}{48} [N]_{-1}^4\cdot \xi^{-2} \eta^{-5}
- \frac{1}{24} [N]_{-2}^2 \cdot \xi^{-3} \eta^{-4} \\
& - & \frac{1}{24} [N]_{-3}^2 \xi^{-4} \eta^{-3}
+ \frac{1}{48} [N]_{-4}^1 \cdot \xi^{-5} \eta^{-2}
+ \frac{1}{48} [N]_{-5}^0 \cdot \xi^{-6} \eta^{-1} + \cdots.
\een
The one-point function is then
\ben
G^{(1)}(\xi) & = &
\sum_{n \geq 1} \frac{(2n-1)!!}{(2n)!} \sum_{p=0}^{2n-1}
(-1)^{p+[(p+1)/2]} \binom{n-1}{[p/2]} \cdot [N]_{-p}^{2n-1-p} \cdot \xi^{-2n-1}.
\een
The following are the first few terms:
\ben
G^{(1)}(\xi)
& = & 1 \cdot N^2 \xi^{-3} + (2 N^3+N)\xi^{-5}
+ (5 N^4 +10 N^2)\xi^{-7} \\
& + & (14 N^5+70N^3+21N)\xi^{-9} \\
& + & (42N^6+420N^4+483N^2) \xi^{-11} \\
& + & (132N^7+2310N^5+6468N^3+1485N) \xi^{-13} \\
& + & (429 N^8+12012N^6+66066N^4+56628N^2) \xi^{-15}+\cdots.
\een
The coefficients are the Harer-Zagier numbers  $\epsilon_g(n)=$number of ways to glue
a $2n$-gon to get a Riemann surface of genus $g$:
\ben
\epsilon_g(n) = \frac{(2n)!}{(n+1)!(n-2g)!} \cdot
\text{coefficient of $x^{2g}$ in \;\;}
\biggl(\frac{x/2}{\tanh(x/2)}\biggr)^{n+1}.
\een
This has been proved by Harer-Zagier \cite{Har-Zag} and Itzykson and Zuber \cite{Itz-Zub}.
Here we present another proof.
Write $G^{(1)}(\xi)= \sum_{n\geq 1} (2n-1)!! \cdot b(n, N) \xi^{-2n-1}$.
We now show that
\be \label{eqn:Rec-b}
b(n, N) = b(n, N-1) +b(n-1, N) + b(n-1, N-1).
\ee
First note:
\ben
&& b(n, N) - b(n, N-1) \\
& = & \frac{1}{(2n)!} \sum_{p=0}^{2n-1}
(-1)^{p+[(p+1)/2]} \binom{n-1}{[p/2]}
\cdot \biggl( [N]_{-p}^{2n-1-p}  - [N-1]_{-p}^{2n-1-p} \biggr)  \\
& = & \frac{1}{(2n)!} \sum_{p=0}^{2n-1}
(-1)^{p+[(p+1)/2]} \binom{n-1}{[p/2]}  \\
&&\cdot [N]_{-p}^{2n-2-p}   \cdot \biggl(
(N+2n-1-p)-(N-1-p)  \biggr) \\
& = & \frac{1}{(2n-1)!} \sum_{p=0}^{2n-1} (-1)^{p+[(p+1)/2]} \binom{n-1}{[p/2]}
\cdot [N]_{-p}^{2n-2-p}.
\een
Now we apply the identity
\ben
[N]_{-p}^{2n-2-p}
= [N]_{-p-1}^{2n-2-p}  + (2n-1) \cdot [N]_{-p}^{2n-3-p}
\een
recursively to get
\be \label{eqn:b-b}
\begin{split}
& b(n,N) - b(n, N-1) \\
= & \frac{1}{(2n-2)!} \sum_{j=0}^{2n-2}
\sum_{p=0}^j (-1)^{p+[(p+1)/2]} \binom{n-1}{[p/2]}
\cdot   [N]_{-j}^{2n-3-j} \\
+ & \frac{1}{(2n-1)!} \sum_{p=0}^{2n-1}
(-1)^{p+[(p+1)/2]} \binom{n-1}{[p/2]}
\cdot [N]_{-(2n-1)}^{-1}.
\end{split}
\ee
We now prove the following identity:
\be \label{eqn:Sum}
\begin{split}
& \sum_{p=0}^j (-1)^p \cdot (-1)^{[(p+1)/2]} \binom{n-1}{[p/2]} \\
= & \begin{cases}
1, & j= 0, \\
(-1)^{j+[(j+1)/2]} \binom{n-2}{[j/2]}
+ (-1)^{j-1+[j/2]} \binom{n-2}{[(j-1)/2]}, & 1 \leq j \leq 2n-2, \\
0, & j = 2n-1.
\end{cases}
\end{split}
\ee
This can be proved as follows.
Denote the left-hand side by $U_j$ and the right-hand side by $V_j$.
It is easy to see that
\begin{align*}
U_0 &=  V_0 = 1, & U_1 & = V_1 = 2, & U_{2n-1} & =V_{2n-1}= 0.
\end{align*}
For $2 \leq j \leq 2n-2$,
we have
\ben
U_j - U_{j-1} & = & (-1)^j \cdot (-1)^{[(j+1)/2]} \binom{n-1}{[j/2]}, \\
V_j - V_{j-1} & = &\biggl[(-1)^j \cdot (-1)^{[(j+1)/2]} \binom{n-2}{[j/2]}
+ (-1)^{j-1} \cdot (-1)^{[j/2]} \binom{n-2}{[(j-1)/2]} \biggr] \\
& - & \biggl[(-1)^{j-1} \cdot (-1)^{[j/2]} \binom{n-2}{[(j-1)/2]}
+ (-1)^{j-2} \cdot (-1)^{[(j-1)/2]} \binom{n-2}{[(j-2)/2]} \biggr] \\
& = & (-1)^j \cdot (-1)^{[(j+1)/2]} \binom{n-2}{[j/2]}
- (-1)^{j} \cdot (-1)^{[(j-1)/2]} \binom{n-2}{[(j-2)/2]} \\
& = & (-1)^{j-2} \cdot (-1)^{[(j-1)/2]} \biggl[
\binom{n-2}{[j/2]} + \binom{n-2}{[j/2]-1} \biggr] \\
& = & (-1)^j \cdot (-1)^{[(j+1)/2]} \binom{n-1}{[j/2]} \\
& = & U_{j} - U_{j-1}.
\een
This completes the proof of \eqref{eqn:Sum}.

On the other hand,
we have
\ben
&& b(n-1, N) + b(n-1, N-1) \\
& = & \frac{1}{(2n-2)!} \sum_{p=0}^{2n-3}
(-1)^p \cdot (-1)^{[(p+1)/2]} \binom{n-2}{[p/2]}
 \cdot \biggl( [N]_{-p}^{2n-3-p}  + [N-1]_{-p}^{2n-3-p} \biggr)  \\
& = & \frac{1}{(2n-2)!} \sum_{p=0}^{2n-3}
(-1)^p \cdot (-1)^{[(p+1)/2]} \binom{n-2}{[p/2]}
 \cdot \biggl( [N]_{-p}^{2n-3-p} + [N]_{-p-1}^{2n-4-p} \biggr)  \\
& = & \frac{1}{(2n-2)!} \sum_{p=0}^{2n-3}
(-1)^p \cdot (-1)^{[(p+1)/2]} \binom{n-2}{[p/2]} \cdot
[N]_{-p}^{2n-3-p}  \\
& + & \frac{1}{(2n-2)!} \sum_{p=1}^{2n-2}
(-1)^{p-1} \cdot (-1)^{[p/2]} \binom{n-2}{[(p-1)/2]} \cdot
[N]_{-p}^{2n-3-p}  \\
& = & b(n, N) - b(n, N-1).
\een
The last equality is just the result of combining \eqref{eqn:b-b}
with \eqref{eqn:Sum}.

In Harer and Zagier \cite{Har-Zag},
the following result on the generating series of $\epsilon_g(n)$ is proved.
Let
\be
C(n, k): =\sum_{0 \leq g \leq n/2} \epsilon_g(n) k^{n+1-2g}.
\ee
Then $C(n,k) = (2n-1)!!c(n,k)$, where $c(n,k)$ is defined by the generating series:
\be
1+2 \sum_{n=0}^\infty c(n,k)x^{n+1}= \biggl(\frac{1+x}{1-x}\biggr)^k,
\ee
or by the recursion relations:
\be \label{eqn:c-Rec}
c(n,k) = c(n,k-1) +c(n-1,k)  +c(n-1,k-1)
\ee
with the boundary conditions $c(0,k)=k$ , $c(n,0) = 0$.
Hence by \eqref{eqn:Rec-b},
$b(n,k) = c(n,k)$.
Therefore,
we have proved the following formula for $c(n, N)$:
\be
c(n, N)  = \frac{1}{(2n)!} \sum_{p=0}^{2n-1}
(-1)^{p+[(p+1)/2]} \binom{n-1}{[p/2]} [N]_{-p}^{2n-1-p}.
\ee
Write $p=2l$ or $p=2l+1$,
\be \label{eqn:c(n,N)}
c(n, N) = \frac{1}{(2n)!} \sum_{l=0}^{n-1} (-1)^l \binom{n-1}{l}
\biggl( [N]_{-2l}^{2n-2l-1}
+[N]_{-(2l+1)}^{2n-2l-2} \biggr).
\ee

By numerical computations we  have found the following identity:
\be \label{eqn:c(n-1,N)}
c(n-1, N) = \frac{1}{(2n)!} \sum_{l=0}^{n-1} (-1)^l \binom{n-1}{l}
\biggl( [N]_{-2l}^{2n-2l-1}
-[N]_{-(2l+1)}^{2n-2l-2} \biggr).
\ee
By \eqref{eqn:c(n,N)} this is equivalent to
\be
c(n, N) - c(n-1,N)
= \frac{2}{(2n)!} \sum_{l=0}^{n-1} (-1)^l \binom{n-1}{l}
 [N]_{-(2l+1)}^{2n-2l-2}.
\ee
By \eqref{eqn:c-Rec},
\ben
&& c(n,N)   \\
& = & \frac{1}{(2n)!} \sum_{l=0}^{n-1} (-1)^l \binom{n-1}{l}
\biggl((N+2n-2l-1) \cdot [N]_{-2l}^{2n-2l-2}
+[N]_{-(2l+1)}^{2n-2l-2} \biggr) \\
& = & \frac{2}{(2n)!} \sum_{l=0}^{n-1} (-1)^l \binom{n-1}{l}
[N]_{-(2l+1)}^{2n-2l-2}  \\
&+& \frac{1}{(2n-1)!} \sum_{l=0}^{n-1} (-1)^l \binom{n-1}{l}
 \cdot [N]_{-2l}^{2n-2l-2}.
\een
So we will prove
\be
c(n-1, N) = \frac{1}{(2n-1)!} \sum_{l=0}^{n-1} (-1)^l \binom{n-1}{l}
 \cdot [N]_{-2l}^{2n-2l-2},
\ee
or by changing $n-1$ to $n$,
\be \label{eqn:c-Simple}
c(n, N) = \frac{1}{(2n+1)!} \sum_{l=0}^{n} (-1)^l \binom{n}{l}
 \cdot [N]_{-2l}^{2n-2l}.
\ee
Write the right-hand as $a(n, N)$. We have
\ben
& & a(n,N) - a(n, N-1) \\
& = & \frac{1}{(2n+1)!} \sum_{l=0}^{n} (-1)^l \binom{n}{l}
 \cdot ([N]_{-2l}^{2n-2l} - [N-1]_{-2l}^{2n-2l}) \\
& = & \frac{1}{(2n)!} \sum_{l=0}^{n} (-1)^l \binom{n}{l}
 \cdot [N]_{-2l}^{2n-2l-1} \\
& = & \frac{1}{(2n)!} [N]_0^{2n-1}+\frac{1}{(2n)!} \sum_{l=1}^n (-1)^l \binom{n}{l}
\cdot [N]_{-2l}^{2n-2l-1} \\
& = & \frac{1}{(2n)!}(2n \cdot [N]_0^{2n-2}+ [N]_{-1}^{2n-2})
+ \frac{1}{(2n)!} \sum_{l=1}^n (-1)^l \binom{n}{l}
\cdot [N]_{-2l}^{2n-2l-1} \\
& = & \frac{1}{(2n)!}(2n \cdot [N]_0^{2n-2} + 2n \cdot [N]_{-1}^{2n-3} + [N]_{-2}^{2n-3}) \\
& + & \frac{1}{(2n)!} \sum_{l=1}^n (-1)^l \binom{n}{l}
\cdot [N]_{-2l}^{2n-2l-1} \\
& = & \frac{1}{(2n-1)!}([N]_0^{2n-2} + [N-1]_{0}^{2n-2}) +
\frac{1}{(2n)!} \biggl(1+(-1)^1 \binom{n}{1}\biggr)[N]_{-2}^{2n-3} \\
& + & \frac{1}{(2n)!} \sum_{l=2}^n (-1)^l \binom{n}{l}
\cdot [N]_{-2l}^{2n-2l-1}.
\een
By repeating this procedures we get:
\ben
& & a(n,N) - a(n, N-1) \\
& = & \frac{1}{(2n-1)!} \sum_{l=0}^{n-1} \sum_{j=0}^l (-1)^j \binom{n}{j} \cdot
\big([N]_{-2l}^{2n-2l-2} +[N-1]_{-2l}^{2n-2l-2} \big) \\
& + & \sum_{j=0}^n (-1)^j \binom{n}{j} \cdot [N]_{-2n}^{-1} \\
& = & \frac{1}{(2n-1)!} \sum_{l=0}^{n-1} (-1)^l \binom{n-1}{l}
 \cdot ([N]_{-2l}^{2n-2l-2} + [N-1]_{-2l}^{2n-2l-2}) \\
& = & a(n-1,N) + a(n-1, N-1).
\een
In the above we have used the following identity for $l=0, \dots, n$:
\be
\sum_{j=0}^l (-1)^j \binom{n}{j} = \binom{n-1}{l}.
\ee
By checking the boundary values one then proves
\eqref{eqn:c-Simple} and hence also \eqref{eqn:c(n-1,N)}.

The $2$-point function is given by
\be
G^{(2)}_N(\xi_1,\xi_2)
= \frac{A(\xi_1, \xi_2) - A(\xi_2,\xi_1)}{\xi_1-\xi_2}
-A(\xi_1, \xi_2)\cdot A(\xi_2,\xi_1).
\ee
After plugging in \eqref{eqn:A},
we get:
\ben
&& G^{(2)}_N(\xi_1,\xi_2) \\
& = & \sum_{n \geq 1} \frac{(2n-1)!!}{(2n)!} \sum_{p=0}^{2n-1}
(-1)^{p+[(p+1)/2]} \binom{n-1}{[p/2]} \cdot [N]_{-p}^{2n-1-p} \\
&& \cdot
\frac{\xi_1^{-p-1}\xi_2^{-2n+p}-\xi_2^{-p-1}\xi_1^{-2n+p}}{\xi_1-\xi_2} \\
& - & \sum_{n \geq 1} \frac{(2n-1)!!}{(2n)!} \sum_{p=0}^{2n-1}
(-1)^{p+[(p+1)/2]} \binom{n-1}{[p/2]}
 \cdot [N]_{-p}^{2n-1-p} \cdot \xi_1^{-p-1} \xi_2^{-2n+p} \\
& \cdot & \sum_{m \geq 1} \frac{(2m-1)!!}{(2m)!} \sum_{q=0}^{2m-1}
(-1)^{q+[(q+1)/2]} \binom{m-1}{[q/2]}
 \cdot [N]_{-q}^{2m-1-q} \cdot \xi_2^{-q-1}
\xi_1^{-2m+q}.
\een
It is interesting to compare with \cite[Example 3.2.5]{Dub-Yan}.
To get some better understanding of this formula,
we consider the coefficients of $\xi_2^{-2}$ and $\xi_2^{-3}$.
The coefficient of $\xi_1^{-2}$ in $G^{(2)}_N(\xi_1,\xi_2)$
is
\ben
&   & \sum_{n \geq 1} \frac{(2n-1)!!}{(2n)!} [N]_{0}^{2n-1} \cdot \xi_2^{-2n}
- \sum_{n \geq 1} (-1)^{n-1}  \frac{(2n-1)!!}{(2n)!}
[N]_{-(2n-1)}^0 \cdot \xi_2^{-2n}  \\
& - & \sum_{k \geq 1} \frac{(2k-1)!!}{(2k)!}
[N]_{0}^{2k-1}  \xi_2^{-2k}
\cdot \sum_{l \geq 1} (-1)^{l-1}  \frac{(2l-1)!!}{(2l)!}
[N]_{-(2l-1)}^{0} \cdot \xi_2^{-2l} \\
& = & \sum_{n \geq 1} \biggl(
\frac{(2n-1)!!}{(2n)!} [N]_{j=0}^{2n-1}
- (-1)^{n-1}  \frac{(2n-1)!!}{(2n)!} [N]_{-(2n-1)}^0  \\
& - & \sum_{\substack{k+l=n\\k,l\geq 1}} \frac{(2k-1)!!}{(2k)!}
[N]_{0}^{2k-1}  \cdot  (-1)^{l-1}  \frac{(2l-1)!!}{(2l)!}
[N]_{-(2l-1)}^{0} \biggr) \cdot \xi_2^{-2n} \\
& = & \sum_{k=0}^\infty \frac{(2k-1)!!}{(2k)!}
[N]_{0}^{2k-1} \cdot \xi_2^{-2k}
\cdot \sum_{l=0}^\infty (-1)^{l}  \frac{(2l-1)!!}{(2l)!}
[N]_{-(2l-1)}^{0}  \cdot \xi_2^{-2l}
-1.
\een
We claim that it is equal to
\be
\sum_{n \geq 1} (2n-1)!! \cdot c(n-1, N) \xi_2^{-2n}
= \sum_{n \geq 1} (2n-1) \cdot C(n-1, N) \xi_2^{-2n},
\ee
in other words,
if we set $(-1)!!=1$ and $c(-1, N) = 1$,
then one has
\be
\begin{split}
& \sum_{n=0}^\infty (2n-1)!! \cdot c(n-1, N) \xi_2^{-2n} \\
= & \sum_{k=0}^\infty \frac{(2k-1)!!}{(2k)!}
[N]_{0}^{2k-1} \cdot \xi_2^{-2k}
\cdot \sum_{l=0}^\infty (-1)^{l}  \frac{(2l-1)!!}{(2l)!}
[N]_{-(2l-1)}^{0}  \cdot \xi_2^{-2l}.
\end{split}
\ee
We need to prove the following identity:
\ben
&& c(n-1, N)
= \frac{1}{(2n)!} \prod_{j=0}^{2n-1} (N+j)
- (-1)^{n-1}  \frac{1}{(2n)!}
\prod_{j=-(2n-1)}^0 (N+j) \\
& - & \frac{1}{(2n-1)!!} \sum_{\substack{k+l=n\\k,l\geq 1}}
\frac{(2k-1)!!}{(2k)!} [N]_{0}^{2k-1} \cdot
(-1)^{l-1}  \frac{(2l-1)!!}{(2l)!} [N]_{-(2l-1)}^{0} \\
& = & \frac{N}{(2n)!} \sum_{k+l=n} (-1)^l\binom{n}{k}\cdot [N]_{-(2l-1)}^{2k-1}.
\een
We use \eqref{eqn:c(n-1,N)} to get:
\ben
c(n-1, N) & = & \frac{1}{(2n)!} \sum_{l=0}^{n-1} (-1)^l \binom{n-1}{l}
\biggl( [N]_{-2l}^{2n-2l-1}
-[N]_{-(2l+1)}^{2n-2l-2} \biggr) \\
& = & \frac{1}{(2n)!} \sum_{l=0}^{n-1} (-1)^l \binom{n-1}{l} [N]_{-2l}^{2n-2l-1} \\
& - & \frac{1}{(2n)!} \sum_{l=0}^{n-1} (-1)^l \binom{n-1}{l}
[N]_{-(2l+1)}^{2n-2l-2} \\
& = & \frac{1}{(2n)!} \sum_{l=0}^{n-1} (-1)^l \binom{n-1}{l} [N]_{-2l}^{2n-2l-1} \\
& + & \frac{1}{(2n)!} \sum_{l=1}^{n} (-1)^l \binom{n-1}{l-1}
[N]_{-(2l-1)}^{2n-2l} \\
& = & \frac{1}{(2n)!} \sum_{l=0}^{n-1} (-1)^l \binom{n-1}{l}(N-2l)\cdot  [N]_{-(2l-1)}^{2n-2l-1} \\
& + & \frac{1}{(2n)!} \sum_{l=1}^{n} (-1)^l \binom{n-1}{l-1}
(N+2n-2l) \cdot [N]_{-(2l-1)}^{2n-2l-1} \\
& = & \frac{N}{(2n)!} \sum_{l=0}^n (-1)^l \binom{n}{l} [N]_{-(l-1)}^{2n-2l-1} \\
& + & \frac{1}{(2n)!} \sum_{l=1}^{n-1} (-1)^l
\biggl(-2l \binom{n-1}{l} + (2n-2l) \binom{n-1}{l-1}\biggr) [N]_{-(l-1)}^{2n-2l-1} \\
& = & \frac{N}{(2n)!} \sum_{l=0}^n (-1)^l \binom{n}{l} [N]_{-(l-1)}^{2n-2l-1}.
\een
Hence we have proved that
\be \label{eqn:c(n-1,N)-Simple}
c(n-1, N) = \frac{N}{(2n)!} \sum_{l=0}^n (-1)^l \binom{n}{l} [N]_{-(l-1)}^{2n-2l-1}.
\ee

The coefficient of $\xi_1^{-3}$ in $G^{(2)}_N$ is
\ben
&& \sum_{n \geq 2} \frac{(2n-1)!!}{(2n)!} \cdot [N]_{0}^{2n-1} \xi_2^{-(2n-1)}
+ \sum_{n \geq 2} \frac{(2n-1)!!}{(2n)!} (-1)^n \cdot [N]_{-(2n-1)}^{0} \xi_2^{-(2n-1)} \\
&+& \sum_{n \geq 2} \frac{(2n-1)!!}{(2n)!} \cdot [N]_{-1}^{2n-2} \xi_2^{-(2n-1)}
+\sum_{n \geq 2} \frac{(2n-1)!!}{(2n)!}(-1)^n \cdot [N]_{-(2n-2)}^{1} \xi_2^{-(2n-1)} \\
& - & \sum_{n \geq 1} \frac{(2n-1)!!}{(2n)!}
  [N]_{-1}^{2n-2} \cdot   \xi_2^{-(2n-1)} \\
& \cdot & \sum_{m \geq 1} \frac{(2m-1)!!}{(2m)!}
(-1)^{m-1}
 \cdot [N]_{-(2m-1)}^{0} \cdot \xi_2^{-2m}  \\
& - & \sum_{n \geq 1} \frac{(2n-1)!!}{(2n)!}
\cdot [N]_{0}^{2n-1} \cdot \xi_2^{-2n} \\
& \cdot & \sum_{m \geq 1} \frac{(2m-1)!!}{(2m)!}
(-1)^{m-1}
 \cdot [N]_{-(2m-2)}^{1} \cdot \xi_2^{-(2m-1)}.
\een
It can be rewritten in the following form:
\ben
&& \sum_{n \geq 2} \frac{1}{n!2^n} \cdot [N]_{0}^{2n-1} \xi_2^{-(2n-1)}
+ \sum_{n \geq 2} \frac{1}{n!2^n} (-1)^n \cdot [N]_{-(2n-1)}^{0} \xi_2^{-(2n-1)} \\
&+& \sum_{n \geq 2} \frac{1}{n!2^n} \cdot [N]_{-1}^{2n-2} \xi_2^{-(2n-1)}
+\sum_{n \geq 2} \frac{1}{n!2^n} (-1)^n \cdot [N]_{-(2n-2)}^{1} \xi_2^{-(2n-1)} \\
& + & \sum_{k,l \geq 1} (-1)^{l} \frac{1}{k!l!2^{k+l}}
 N(N-1)\cdot [N]_{-(2l-1)}^{2k-2} \cdot   \xi_2^{-(2(k+l)-1)}   \\
& + & \sum_{k,l \geq 1} (-1)^l\frac{1}{k!l!2^{k+l}}
\cdot  N(N+1) \cdot [N]_{-(2l-2)}^{2k-1} \cdot \xi_2^{-(2(k+l)-1)} \\
& = & N(N-1) \sum_{n \geq 2} \frac{(2n-1)!!}{(2n)!}
\sum_{l=0}^n (-1)^l \binom{n}{l} [N]_{-(2l-1)}^{2n-2l-2} \cdot \xi_2^{-(2n-1)} \\
& + & N(N+1) \sum_{n \geq 2} \frac{(2n-1)!!}{(2n)!}
\sum_{l=0}^n (-1)^l \binom{n}{l} [N]_{-(2l-2)}^{2n-2l-1} \cdot \xi_2^{-(2n-1)}.
\een
We claim that it is equal to
\be
\sum_{n \geq 2} 2(n-1) \cdot (2n-3)!! \cdot c(n-1, N) \xi_2^{-(2n-1)}
= \sum_{n \geq 2} 2(n-1)\cdot C(n-1, N) \xi_2^{-(2n-1)}.
\ee
By \eqref{eqn:c(n-1,N)-Simple} we need to show that
\ben
&& N(N-1)  \frac{(2n-1)!!}{(2n)!}
\sum_{l=0}^n (-1)^l \binom{n}{l} [N]_{-(2l-1)}^{2n-2l-2}   \\
& + & N(N+1)  \frac{(2n-1)!!}{(2n)!}
\sum_{l=0}^n (-1)^l \binom{n}{l} [N]_{-(2l-2)}^{2n-2l-1}   \\
& = &  2(n-1) \cdot (2n-3)!! \cdot c(n-1, N)  \\
& = &  2(n-1) \cdot (2n-3)!! \cdot \frac{N}{(2n)!}
\sum_{k+l=n} (-1)^l\binom{n}{k}\cdot [N]_{-(2l-1)}^{2k-1},
\een
or equivalently,
\ben
&&    (2n-1) \cdot (N-1)
\sum_{l=0}^n (-1)^l \binom{n}{l} [N]_{-(2l-1)}^{2n-2l-2}   \\
& + &  (2n-1) \cdot (N+1)
\sum_{l=0}^n (-1)^l \binom{n}{l} [N]_{-(2l-2)}^{2n-2l-1}   \\
& = &  2(n-1) \cdot   \sum_{l=0}^n (-1)^l\binom{n}{l} [N]_{-(2l-1)}^{2n-2l-1}.
\een
Denote by $LHS$ and $RHS$ the left-hand side and the right-hand side respectively.
Then we have
\ben
LHS-RHS  &= & \sum_{l=0}^n (-1)^l\binom{n}{l}
(2nN^2 + (2n-4l)N + 8(n-1) l(n-l) ) [N]_{-(2l-2)}^{2n-2l-2}.
\een
This is easily shown to vanish by checking
that for $0 \leq l \leq n$,
\ben
&& (-1)^l\binom{n}{l}
(2nN^2 + (2n-4l)N + 8(n-1) l(n-l) ) [N]_{-(2l-2)}^{2n-2l-2} \\
& = & 2n \cdot (-1)^l \binom{n-1}{l} [N]_{-2l}^{2n-2-2l}
- 2n \cdot (-1)^{l-1} \binom{n-1}{l-1} [N]_{-2(l-1)}^{2n-2-2(l-1)}.
\een
From this we also get:
\ben
&& \sum_{l=0}^m (-1)^l\binom{n}{l}
(2nN^2 + (2n-4l)N + 8(n-1) l(n-l) ) [N]_{-(2l-2)}^{2n-2l-2} \\
& = &  2n \cdot (-1)^m \binom{n-1}{m} [N]_{-2m}^{2n-2-2m}.
\een

To summarize,
we have shown a surprising connection between the one-point function and the two-point
function of the Hermitian one-matrix model:
\be
\begin{split}
G_N^{(2)}(\xi_1,\xi_2)
= & \xi_1^{-2} \sum_{n \geq 1} (2n-1) \cdot C(n-1, N) \xi_2^{-2n} \\
+ & \xi_1^{-3} \sum_{n \geq 2} (2n-2) \cdot C(n-1, N) \xi_2^{-(2n-1)} +\cdots,
\end{split}
\ee
where $G_N^{(1)}(\xi)  =  \sum_{n=1}^\infty C(n,N) \xi^{-n-1}$
is the one-point function computed by Harer and Zagier \cite{Har-Zag}.
We expect that further investigations of more terms in $G_N^{(2)}$
and of $G_N^{(n)}$ for $n >2$ will reveal more clearly the relationship to $G^{(1)}_N$.

\begin{remark}
There are similar but different formulas for $c(n, N)$ in the literature.
For example,
\cite[(6.5.22)]{Mehta} reads:
\be
c(n,N) = \sum_{j=0}^n \binom{n}{j} \binom{N}{j+1}2^j
= \sum_{j=0}^n \binom{n}{j} \frac{2^j}{(j+1)!} \cdot [N]_{-j}^0£¬
\ee
and \cite[(6.5.29)]{Mehta} reads:
\be
c(n,N) = \half \sum_{j_1+j_2=n+1} \binom{N}{j_1} \binom{N+j_2-1}{j_2}
= \half N \sum_{j_1+j_2=n+1} \frac{1}{j_1!j_2!} [N]_{-j_1+1}^{j_2-1}.
\ee

\end{remark}

\subsection{A family of $\tau$-functions of the KP hierarchy by Hermitian matrix model}
\label{sec:tHooft}

In the above we have seen that Hermitian one-matrix model defined
by formal Gaussian integrals on the space $N\times N$ Hermitian matrices
defines a tau-function $Z_N$ of the KP hierarchy,
and we have presented a formula that computes
the $n$-point functions
as a formal power series whose coefficients are polynomials in $N$
with nonnegative integers as coefficients.
Even though we have $N$ as a fixed positive integer to start with,
by now we can treat it as a parameter that can take any real value.
Hence by changing $N$ to $t$,
we get a family $\{Z_t\}_{t\in \bR}$ of $\tau$-functions of the KP hierarchy,
whose associated $n$-point functions can be obtained by replacing $N$
by $t$ in all our formulas.
This can be achieved in the framework of Hermitian matrix model
by introducing the 't Hooft coupling constant
\be
t = N g_s.
\ee
In other words, one can take the coupling constant $g_s$ to be
\be
g_s = \frac{1}{N} t,
\ee
and the formal matrix integral is then changed to:
\be
Z_N = \frac{\int_{\bH_N} dM \exp \biggl( N \tr
\sum\limits_{n=1}^\infty \frac{g_n-\delta_{n,2}}{nt} M^n\biggr)}
{\int_{\bH_N} dM \exp \biggl( -\frac{N}{2t} \tr (M^2)\biggr)}.
\ee
The correlators for this model can be computed using the following change of variable:
\be
N = g_s^{-1}t.
\ee
For example,
in degree two we have
\ben
&& \corr{p_2}_N = N^2 = t^2g_s^{-2} ,\\
&& \corr{p_1^2}_N = Ng_s^{-1} = tg_s^{-2},
\een
in degree four,
\ben
&& \corr{p_4}_N = (N + 2 N^3)g_s = t + 2t^3 g_s^{-2}, \\
&& \corr{p_3p_1}_N = 3N^2 = 3 t^2g_s^{-2}, \\
&& \corr{p_2^2}_N =  2N^2 + N^4 = 2t^2 g_s^{-2} + t^4 g_s^{-4} , \\
&& \corr{p_2p_1^2}_N = (2N + N^3)g_s^{-1} = 2t g_s^{-2} + t^3 g_s^2, \\
&& \corr{p_1^4}_N = 3N^2g_s^{-2} = 3t^2 g_s^{-4},
\een
and in degree six:
\ben
&& \corr{p_6}_N = (10 N^2 + 5 N^4)g_s^2 = 10t^2 + 5t^4g_s^{-2}, \\
&& \corr{p_5p_1}_N = (5N + 10N^3)g_s = 5t +10 t^3 g_s^{-2}, \\
&& \corr{p_4p_2}_N = (4N + 9 N^3 + 2N^5) g_s = 4t +9t^3g_s^{-2} + 2t^5g_s^{-4}, \\
&& \corr{p_4p_1^2}_N = 13N^2 + 2N^4 =13t^2g_s^{-2} + 2t^4 g_s^{-4}, \\
&& \corr{p_3^2}_N = (3N + 12N^3)g_s = 3t +12 t^3 g_s^{-2}, \\
&& \corr{p_3p_2p_1}_N = 12N^2 + 3N^4= 12t^2g_s^{-2} + 3t^4g_s^{-4}, \\
&& \corr{p_3p_1^3}_N = (6N+9N^3) g_s^{-1} = 6tg_s^{-2}+9t^3g_s^{-4}, \\
&& \corr{p_2^3}_N = 8N^2+6N^4+N^6 = 8t^2g_s^{-2}+6t^4g_s^{-4}+t^6g_s^{-6} , \\
&& \corr{p_2^2p_1^2}_N = (8N+6N^3+ N^5)g_s^{-1} = 8tg_s^{-2}+6t^3g_s^{-4}+ t^5g_s^{-6}, \\
&& \corr{p_2p_1^4}_N = (12N^2 + 3N^4) g_s^{-2} = 12t^2 g_s^{-4} +3 t^4 g_s^{-6}, \\
&& \corr{p_1^6}_N = 15 N^3g_s^{-3} = 15t^3 g_s^{-6}.
\een

By \eqref{eqn:correlator},
\be
\corr{\frac{1}{z_\lambda} p_\lambda}_N^c
= \sum_{\Gamma \in \Gamma^{\lambda c}}
\frac{1}{|\Aut(\Gamma)|} g_s^{\half|\lambda| - l(\lambda)-|F(\Gamma)|} t^{|F(\Gamma)|},
\ee
Denote by $\Sigma_\Gamma$ the closed surface obtained from
the fat graph $\Gamma$ by
filling $|F(\Gamma)|$ discs along the boundary components.
Then one has
\be
\half |\lambda| - l(\lambda) - |F(\lambda)|
= - \chi(\Sigma_\Gamma) = 2g(\Sigma_\Gamma)- 2,
\ee
where $g(\Sigma_\Gamma)$ is the genus of $\Sigma_\Gamma$.
\be
\corr{\frac{1}{z_\lambda} p_\lambda}_N
= \sum_{\Gamma \in \Gamma^{\lambda }}
\frac{1}{|\Aut(\Gamma)|} g_s^{-\chi(\Sigma_\Gamma)} \cdot t^{|F(\Gamma)|}.
\ee
An amazing fact is that the right-hand side is now independent of $N$,
it counts graphs on {\em closed} surfaces.
Write
\be
Z_t:=\sum_\lambda \sum_{\Gamma \in \Gamma^{\lambda }}
\frac{1}{|\Aut(\Gamma)|} g_s^{-\chi(\Sigma_\Gamma)} \cdot t^{|F(\Gamma)|}
\cdot p_\lambda,
\ee
then we have
\be
Z_t|_{g_s=1} = Z_N|_{g_s =1, N \to t}.
\ee
In particular, $Z_t|_{g_s=1}$ gives us a family of $\tau$-functions of
the KP hierarchy.
After replacing $N$ by $t$,
the results in Theorem \ref{thm:Main} and Theorem \ref{thm:Main2} also hold
for $Z_t$.

\vspace{.2in}
{\bf Acknowledgements}.
The author is partly supported by NSFC grant 11661131005.
The results in this paper were obtained when the author was preparing for a talk
at Russian-China Conference on Integrable
Systems and Geometry, held at Euler International Mathematical Institute,
St. Petersburg. The author thanks the organizers
and the participants for the hospitality
enjoyed at this conference.

\newpage
\begin{appendices}
\section{Examples of $n$-Point Functions of Hermitian One-Matrix Models}

In this Appendix we present some concrete examples
of $n$-points functions of Hermitian one-matrix models 
computed by the formula in Theorem \ref{thm:Main2}.

\subsection{One-point function}

For $n=1$,
\be
G_N^{(1)}(x) = A(x, x).
\ee
The following are the first few terms:
\ben
 G_N^{(1)}(x)
&= & N^2x^{-3}+(2N^3+N)x^{-5}+(5N^4+10N^2)x^{-7}\\
&+& (14N^5+70N^3+21N)x^{-9}
+(42N^6+420N^4+483N^2)x^{-11} \\
&+&(132N^7+2310N^5+6468N^3+1485N)x^{-13} \\
&+& (429N^8+12012N^6+66066N^4+56628N^2)x^{-15} \\
&+& (1430N^9+60060N^7+570570N^5+1169740N^3+225225N)x^{-17} \\
&+& (4862N^{10}+291720N^8+4390386N^6+17454580N^4 \\
&& +12317877N^2)x^{-19} \\
&+& (16796N^{11}+1385670N^9+31039008N^7+211083730N^5\\
&& +351683046N^3+59520825N)x^{-21} +\cdots.
\een
See A035309 of The On-Line Encyclopedia of Integer Sequences for more terms.

\subsection{Two-point function}
For $n=2$, 
\be
G_N^{(2)}(x,y) = \frac{A(x,y)-A(y,x)}{x-y}- A(x,y) \cdot A(y,x).
\ee
the following are the next few terms:
\ben
 G_N^{(2)}(x,y)
&= & Nx^{-2}y^{-2} + 3N^2(x^{-4}y^{-2}+y^{-2}x^{-4})+2N^2x^{-3}y^{-3}  \\
& + & (10N^3+5N)(x^{-6}y^{-2}+x^{-2}y^{-6} ) \\
&+ & (8N^3+4N)(x^{-5}y^{-3} + x^{-3}y^{-5}) \\
& + & (12N^3+3N)x^{-4}y^{-4} + (35N^4+70N^2)(x^{-8}y^{-2}+x^{-2}y^{-8}) \\
& + & (30N^4+60N^2)(x^{-7}y^{-3}+ x^{-3}y^{-7}) \\
& + & (45N^4+60N^2) (x^{-6}y^{-4}+x^{-4}y^{-6}) \\
& + & (36N^4+60N^2)x^{-5}y^{-5}\\
& + & (126N^5+630N^3+189N)(x^{-10}y^{-2}+x^{-2}y^{-10}) \\
& + & (112N^5+560N^3+168N)(x^{-9}y^{-3}+x^{-3}y^{-9})\\
& + & (168N^5+630N^3+147N)(x^{-8}y^{-4}+x^{-4}y^{-8}) \\
& + & (144N^5+600N^3+156N)(x^{-7}y^{-5}+x^{-5}y^{-7}) \\
& + & (180N^5+600N^3+165N)x^{-6}y^{-6}
\een
the next few terms are
\ben
& + & (462N^6+4620N^4+5313N^2)(x^{-12}y^{-2}+x^{-2}y^{-12})\\
& + & (420N^6+4200N^4+4830N^2)(x^{-11}y^{-3}+x^{-3}y^{-11}) \\
& + & (630N^6+5040N^4+4725N^2)(x^{-10}y^{-4}+x^{-4}y^{-10}) \\
& + & (560N^6+4760N^4+4760)(x^{-9}y^{-5}+x^{-5}y^{-9}) \\
& + & (700N^6+4900N^4+4795N^2)(x^{-8}y^{-6}+x^{-6}y^{-8}) \\
& + & (600N^6+4800N^4+4770N^2)x^{-7}y^{-7}
\een
and the next few terms are
\ben
& +&(1716N^7+30030N^5+84084N^3+19305N)(x^{-14}y^{-2}+x^{-2}y^{-14}) \\
&+& (77616N^3+1584N^7+27720N^5+17820N)(x^{-13}y^{-3}+x^{-3}y^{-13}) \\
&+&( 2376N^7+34650N^5+81774N^3+16335N)(x^{-12}y^{-4}+x^{-4}y^{-12}) \\
&+&( 2160N^7 +32760N^5 +80640N^3 +16740N)(x^{-11}y^{-5}+x^{-5}y^{-11}) \\
& + & ( 2700N^7 +34650N^5 +80640N^3 +17145N)(x^{-10}y^{-6}+x^{-6}y^{-10})\\
&+& ( 2400N^7 +33600N^5 +80640N^3 +16920N)(x^{-9}y^{-7}+x^{-7}y^{-9}) \\
&+&(2800N^7 +34300N^5 +81340N^3 +16695N)(x^{-8}y^{-8})
\een
and the next few terms are
\ben
& + & ( 6435N^8 +180180N^6+990990N^4
+849420N^2) (x^{-16}y^{-2}+x^{-2}y^{-16}) \\
& +& ( 6006N^8+168168N^6+924924N^4
+ 792792N^2 ) (x^{-15}y^{-3}+x^{-3}y^{-15}) \\
& +& ( 9009N^8 +216216N^6+1027026N^4
+ 774774N^2 )(x^{-14}y^{-4}+x^{-4}y^{-14}) \\
& + & (  8316N^8 +205128N^6+1003464N^4
+ 778932N^2 )(x^{-13}y^{-5}+x^{-5}y^{-13})\\
& +& (   10395N^8 +221760N^6+1011780N^4
+783090N^2)(x^{-12}y^{-6}+x^{-6}y^{-12})) \\
& + & (  9450N^8 +214200N^6+1008000N^4
+781200N^2 )(x^{-11}y^{-7}+x^{-7}y^{-11}) \\
& + & ( 11025N^8 +220500N^6+1014300N^4
+781200N^2 )(x^{-10}y^{-8}+x^{-8}y^{-10}) \\
& + & ( 9800N^8 +215600N^6+1009400N^4
+781200N^2 )x^{-9}y^{-9}
\een
and the next few terms are
\ben
&+&(24310N^9+1021020N^7 +9699690N^5
+19885580N^3+3828825N)\\&&\cdot (x^{-18}y^{-2}+x^{-2}y^{-18})\\
& +& (22880N^9 +960960N^7 +9129120N^5
+18715840N^3+3603600N)\\&&\cdot(x^{-17}y^{-3}+x^{-3}y^{-17})  \\
& +& (34320N^9 +1261260N^7 +10540530N^5
+19244940N^3+3378375N)\\&&\cdot (x^{-16}y^{-4}+x^{-4}y^{-16}) \\
& +& (32032N^9 +1201200N^7 +10258248N^5
+19139120N^3 +3423420N)\\&&\cdot(x^{-15}y^{-5}+x^{-5}y^{-15}) \\
& +& (40040N^9 +1321320N^7 +10480470N^5
+19149130N^3 +3468465N) \\&&\cdot(x^{-14}y^{-6}+x^{-6}y^{-14})\\
&+&(36960N^9+1275120N^7 +10395000N^5 +19145280N^3
+3451140N )\\&&\cdot(x^{-13}y^{-7}+x^{-7}y^{-13}) \\
&+&(43120N^9+1325940N^7+10461990N^5+19194560N^3
+3433815N)\\&&\cdot(x^{-12}y^{-8}+x^{-8}y^{-12}) \\
& +& (39200N^9+1293600N^7 + 10419360N^5+19163200N^3+3444840N)
\\&&\cdot (x^{-11}y^{-9}+x^{-9}y^{-11} ) \\
&+&(44100N^9+1323000N^7+10478160N^5+19158300N^3
+3455865N )\\
&&\cdot x^{-10}y^{-10}+ \cdots
\een

\subsection{Two-point function}
For $n=3$, 
\ben
G_N^{(3)}(x,y,z) & = & (\frac{1}{x-y}+A(x,y))(\frac{1}{y-z}+A(y,z))(\frac{1}{z-x}+A(z,x)) \\
&+& (\frac{1}{y-x}+A(y,x))(\frac{1}{z-y}+A(z,y))(\frac{1}{x-z}+A(x,z)).
\een
The first few terms of $G^{(3)}$ are
\ben
G_N^{(3)}(x,y,z)&=& 2N\biggl(\frac{1}{x^{3}y^{2}z^{2}} +\frac{1}{x^{2}y^{3}z^{2}}
+ \frac{1}{x^{2}y^{2}z^{3}} \biggr) \\
& + & 12N^2\biggl(\frac{1}{x^5y^2z^2} + \frac{1}{x^2y^5z^2}+\frac{1}{x^2y^2z^5} \biggr) \\
& + & 12N^2\biggl(\frac{1}{x^4y^3z^2}+\frac{1}{x^4y^2z^3}+\frac{1}{x^3y^4z^2}
+ \frac{1}{x^3y^2z^4}+\frac{1}{x^2y^3z^4}+\frac{1}{x^2y^4z^3}\biggr) \\
& + & \frac{8N^2}{x^3y^3z^3} +\cdots.
\een
To simplify notations,
we write it in the following form:
\ben
G_N^{(3)} & = & 2N[3,2,2]+12N^2[5,2,2]+ 12N^2[4,3,2]+ 8N^2 [3,3,3] +\cdots.
\een
the next few terms are
\ben
&+& (60N^3+30N)[7,2,2]+(60N^3+30N)[6,3,2]+(72N^3+24N)[5,4,2] \\
&+& (48N^3+24N)[5,3,3]+(72N^3+24N)[4,4,3]
\een

\ben
&+& (280N^4+560N^2)[9,2,2]+(280N^4+560N^2)[8,3,2] \\
&+& (360N^4+540N^2)[7,4,2] +(240N^4+480N^2)[7,3,3] \\
&+& (360N^4+540N^2)[6,5,2]+(360N^4+480N^2)[6,4,3] \\
&+& (288N^4+480N^2)[5,5,3] + (432N^4+468N^2) [5,4,4].
\een

\ben
& + & (1260N^5+6300N^3+1890N)[11,2,2]+(1260N^5+6300N^3+1890N)[10,3,2]\\
& + & (1680N^5+6720N^3+1680N)[9,4,2] +(1120N^5+5600N^3+1680N)[9,3,3] \\
&+& (1680N^5+6720N^3+1680N)[8,5,2]+(1680N^5+6300N^3+1470N)[8,4,3] \\
&+& (1800N^5+6600N^3+1770N)[7,6,2] + (1440N^5+6000N^3+1560N) [7,5,3] \\
&+& (2160N^5+6660N^3+1350N)[7,4,4] + (1800N^5+6000N^3+1650N) [6,6,3] \\
&+& (2160N^5+6480N^3+1440N)[6,5,4] + (1728N^5+6336N^3+1440N)[5,5,5]
\een

\end{appendices}

\end{document}